\documentclass[manuscript]{emulateapj}
\newcommand{\hei}{He~\textsc{i}}
\newcommand{\heii}{He~\textsc{ii}}

\newcommand{\teff}{T_{\rm eff}}

\shorttitle{Oe Stars in the SMC}
\shortauthors{J. Golden-Marx et al. }

\slugcomment{*** Accepted in ApJ ***}

\begin{document}
\title{Classical O\MakeLowercase{e} Stars in the Field of the Small
  Magellanic Cloud\footnotemark[*]}
\footnotetext[*]{This paper includes data gathered with the 6.5 meter
  Magellan Telescopes located at Las Campanas Observatory, Chile.}

\author{Jesse B.\ Golden-Marx\altaffilmark{1}, M. S.\ Oey\altaffilmark{1}, J. B.\ Lamb\altaffilmark{2}, Andrew S.\ Graus\altaffilmark{3}, Aaron S.\ White\altaffilmark{1}}

\altaffiltext{1}{Department of Astronomy, University of Michigan, 1085 S. University Ave, Ann Arbor, MI, 48109-1107; jessegm@umich.edu}
\altaffiltext{2}{Department of Physical Sciences, Nassau Community College, One Education Drive Garden City, NY, 11530-6793}
\altaffiltext{3}{Department of Physics and Astronomy, University of California at Irvine, Irvine, CA, 92697-4575}

\begin{abstract}
We present $29\pm1$ classical Oe stars from RIOTS4, a spatially complete, spectroscopic survey of Small Magellanic Cloud (SMC) field OB stars.  The two earliest are O6e stars, and four are earlier than any Milky Way (MW) Oe stars.  We also find ten Ope stars, showing \hei\ infill and/or emission; five appear to be at least as hot as $\sim$O7.5e stars.  The hottest, star 77616, shows \heii\ disk emission, suggesting that even the hottest O stars can form decretion disks, and offers observational support for theoretical predictions that the hottest, fastest rotators can generate He$^+$-ionizing atmospheres.  Our data also demonstrate that Ope stars correspond to Oe stars earlier than O7.5e with strong disk emission.  We find that in the SMC, Oe stars extend to earlier spectral types than in the MW, and our SMC Oe/O frequency, $0.26\pm0.04$, is much greater than the MW value, $0.03\pm0.01$.  These results are consistent with angular momentum transport by stronger winds suppressing decretion disk formation at higher metallicity.  In addition, our SMC field Oe star frequency is indistinguishable from that for clusters, which is consistent with the similarity between rotation rates in these environments, and contrary to the pattern for MW rotation rates. Thus, our findings strongly support the viscous decretion disk model and confirm that Oe stars are the high-mass extension of the Be phenomenon.  Additionally, we find that Fe~\textsc{ii} emission occurs among Oe stars later than O7.5e with massive disks, and we revise a photometric criterion for identifying Oe stars to $J-[3.6] \geq 0.1$. 
\end{abstract}


\section{Introduction} 
\label{s:intro}

Despite their scarcity and short lifespans, massive stars shape the evolution of their host galaxies through strong stellar winds and radiative feedback, which shock heat and ionize the surrounding gas, injecting energy into the interstellar medium (ISM).  Moreover, when massive stars end their lives as core-collapse supernovae, the shockwaves can trigger or suppress star formation while enriching the ISM with metals.  Thus, distinct types of feedback are associated with different evolutionary phases of massive stars.  Understanding massive star evolution is complex due in part to the effects of mass loss and the difficulty of modeling their atmospheres, which cannot be treated as being in LTE.  Recent stellar evolution models show that properties such as rotation and binarity also strongly affect massive star evolution.  According to models by \citet{eks12}, rapid rotation induces rotational mixing, which increases the time these stars spend on the main sequence by continually replenishing their cores with H.  Thus, rapid rotation increases the main sequence lifetimes of massive stars and allows them to emit more H-ionizing photons at later times in their evolution \citep[e.g.,][]{lev12}.  Binarity is also key, since observations suggest that the majority of massive stars are binaries \citep[e.g.,][]{san14}, and \citet{dem13} suggest that binary interactions increase the rotation rates of $\sim$20\% of massive stars.  

Understanding different types of evolved massive stars may offer clues about the physical processes that drive their evolution.  Classical Be stars are apparently a subclass of somewhat evolved massive stars whose formation depends on rapid rotation and possibly on binarity.  Today, classical Be stars are characterized as rapidly rotating, non-supergiant B stars with spectra containing Balmer emission lines, which are often double peaked \citep[e.g.,][]{col87,riv13}, resulting from Keplerian rotation of a circumstellar disk \citep[e.g.,][]{str31}.  In addition to the Balmer emission, Be star spectra can show three types of variability \citep{mcl61}: Balmer line variability; variation in the ratio of the violet and red components of the double-peaked Balmer emission lines, or V/R variability; and the appearance or disappearance of a narrow Balmer absorption core, known as a shell absorption spectrum \citep[e.g.,][]{han96}.    

\citet{str31} presented the first viable Be star model, which proposed that the circumstellar disks surrounding Be stars form when the stellar rotation rate $v_{\rm rot}$ approaches the critical velocity $v_{\rm crit}$, where the gravitational and centrifugal forces are equal, causing the centrifugal force to eject the equatorial layer into a disk.  Struve's model omits stars whose spectra contain P~Cygni profiles because Balmer emission from classical Be stars originates from a circumstellar disk, not an isotropic wind, emphasizing that rapid rotation is the primary cause of the Be phenomenon.

Although Struve's model served as the foundation for understanding Be stars, observations suggest that on average, Be stars rotate at only $\sim$0.80$~v_{\rm crit}$ \citep[e.g.,][]{riv06, mei12, cha01}.  However, gravitational limb darkening may cause these rotational rates to be underestimated \citep[e.g.,][]{sto68, tow04, fre05}, thus suggesting that Be stars may rotate closer to $v_{\rm crit}$ than observations find.  Determining how Be stars form with subcritical rotation rates led to the viscous decretion disk model \citep[VDD;][]{lee91}.  This model proposes that processes such as non-radial pulsation (NRP) spin up the stellar atmosphere to slightly super-Keplerian rotation speeds and that if the outer layers are continually supplied with angular momentum, they are lifted from the star and move into a disk \citep[e.g.,][]{lee91, por99}.  These viscous disks are characterized by outward angular momentum transfer due to a turbulent viscosity and resemble models for accretion disks, but with angular momentum supplied at the inner boundary of the disk \citep{lee91}.  Interferometric observations and observations of V/R and photometric variability from the disk support the VDD model and its predictions \citep[e.g.,][]{riv13}.  

Currently, NRPs are the favored mechanism for ejecting sub-critically rotating circumstellar material into a Keplerian orbit.  NRPs require $\Delta v = v_{\rm crit} - v_{\rm rot}$ to be approximately equal to the pulsation amplitude, which for NRPs can approach or slightly exceed the sound speed \citep{owo06}.  Therefore, NRPs require the star to rotate close to $v_{\rm crit}$.  However,  for some Be stars with slower $v_{\rm rot}$, NRPs may be unable to eject the material and form decretion disks, so the mechanism for transporting material to the disk remains an open question.      

Regardless of ejection mechanism, the VDD model requires Be stars rotate near $v_{\rm crit}$ to form decretion disks.  Stellar evolution models demonstrate that as isolated stars evolve, the surface rotation velocity, $v_{\rm rot}$, can approach $v_{\rm crit}$, even though $v_{\rm rot}$ decreases \citep[e.g.,][]{lan98, mey03}.  The rotation rate, $\Omega$, also evolves towards the Keplerian limit if the star retains angular momentum while on the main sequence and transports it efficiently between its core and envelope \citep[e.g.,][]{neg04, eks08, dem13}.  Additionally, binary interactions have been proposed to spin-up Be stars to $v_{\rm crit}$.  \citet{mcs05} suggest that binary interactions may be involved in spinning up as many as 73\% of Be stars; however much debate exists about the role of binarity in forming Be stars.   

One way to evaluate the VDD model is to study Oe stars, first identified by \citet{con74} and proposed as an early-type extension of the Be phenomenon due to He~\textsc{ii} absorption and Balmer emission.  \cite{con74} defined Oe stars as rapidly rotating O stars with spectra containing Balmer emission, but not the characteristic Of star emission features, He~\textsc{ii}~$\lambda$4686 and N~\textsc{iii}~$\lambda\lambda$4634-40-42, associated with strong stellar winds.  Oe stars also show additional characteristics similar to classical Be stars, including V/R and Balmer line variability \citep{div83, rau07} and in some cases, He~\textsc{i} emission \citep[e.g.,][]{fro76}.  However, polarization measurements from \citet{vin09} have not been able to confirm that circumstellar disks surround Oe stars.  Thus, there is still some uncertainty that Oe stars are the high-mass analogs of the Be phenomenon.    

There is another rare class of O stars which has also been proposed to correspond to higher-mass analogs of the Be phenomenon, the Onfp \citep{wal73} or Oef \citep{con74} stars.  These are O stars with double-peaked He~\textsc{ii}~$\lambda$4686 emission. However, unlike Oe stars, Onfp stars show Balmer emission only in H$\alpha$ \citep{con74}.  Furthermore, while double-peaked He~\textsc{ii}~$\lambda$4686 emission may suggest a disk, more recently, this emission has been proposed to come from stellar winds \citep[e.g.,][]{wal10}.  The status of Onfp stars remains uncertain, and Oe stars are generally considered to be the early-type analogs to Be stars.  However, there have been few studies of Oe stars as a class.  The study with the largest sample to date is by \citet{neg04}, who took new spectra of all previously identified Oe, emission-line O, and peculiar O stars in the Milky Way (MW) visible from La Palma Observatory.

O stars have stronger winds than B stars, so the frequency of Oe relative to Be stars should provide a critical probe of the VDD model.  Since stellar wind strength increases with stellar mass and metallicity, in high-metallicity environments, line-driven winds from higher-mass stars should remove angular momentum and inhibit rotation near $v_{\rm crit}$, suppressing decretion disk formation among early-type stars.  \citet{neg04} stress that the Galactic frequency of Oe stars is low, only $0.04\pm0.02$ among O7.5 -- O9 stars, based on the catalog of \citet{mai04}.  We confirm an Oe/O ratio of $0.03\pm0.01$, where the denominator includes Oe and O stars, based on luminosity class III-V stars from the Galactic O Star Spectroscopic Survey \citep[GOSSS;][]{sot11, sot14, sot14b}, a complete magnitude-limited survey of Galactic O stars with $B < 8$. We caution this Oe/O value may be underestimated if stars with weak Balmer emission are not identified; however, even if twice as many MW Oe stars exist, the frequency is far less than the observed Be/B ratio, $0.17\pm 0.03$\footnote{\footnotesize{No uncertainty is provided for this value. Based on the quoted errors for the frequencies of individual spectral types, the aggregate uncertainty is likely $\lesssim 0.03$.  We henceforth adopt this value.}} \citep{zor97}, for B stars in the MW field.  Moreover, the change in frequency of the Be phenomenon between O and B stars is steep.  The Be/B ratio is $0.27\pm0.01$ for B0 stars in the MW field \citep{zor97}, far greater than $0.05\pm0.03$, the frequency for O9.5 stars in the MW \citep{sot14b}.  Thus, these frequencies are qualitatively consistent with the expected effect of wind strength on decretion disk formation.  \citet{neg04} point out that since O8--9~V stars are the progenitor population of B0--1~IIIe stars, the high frequency of Be stars among B0--1~III stars implies that $\Omega/~\Omega_{\rm crit}$ must increase during the main sequence phase; $\Omega_{\rm crit}$, the critical rotation rate, has a non-linear relation to $v_{\rm crit}$ \citep[e.g.][]{cha01, mar06b, riv13}. 

Metallicity, as well as stellar mass, affects stellar wind strength and the ability to form decretion disks.  Existing data support this interpretation; in the low-metallicity Small Magellanic Cloud (SMC), on average, B stars rotate at 0.58~$\Omega_{\rm crit}$, which is much faster than in the MW, where B stars rotate at 0.30--0.40~$\Omega_{\rm crit}$ \citep{mar07}.  These observations are consistent with findings from \citet{mar10} that the frequency of Oe/Be stars among early-type stars in the SMC is $\sim$3--5 times higher than in the MW.  Measurements of the Be/B ratio across all B spectral types are estimated to be between 0.20--0.40 in the SMC \citep{mar07b}, while \citet{zor97} obtain a frequency of $0.17\pm0.03$ in the MW field.  In low-metallicity systems, the frequency of the Be phenomenon should also be enhanced among earlier-type OB stars.  Reports of an SMC O7~Ve \citep{mas95}, an SMC O4--7e \citep{eva04} star, and an O3e star \citep{con86} in the Large Magellanic Cloud (LMC) suggest that this may indeed be the case, since stars of such early types have not been found in the Galaxy (\S~\ref{s:earlyOe}). 

Additionally, metallicity impacts NRPs.  Since the pulsational amplitude is approximately equal to $\Delta v$, in low-metallicity environments, where OB stars rotate faster, it should be easier for NRPs to drive decretion disk formation.  Moreover, rapid rotation leads to rotational mixing, which enriches the metallicity in the outer layers of stars, enhancing NRPs \citep{mae01}.  Thus, the faster rotation in low-metallicity environments favors pulsation, explains the higher frequency of pulsating Be stars in low-metallicity environments \citep{dia09}, and supports the anticorrelation between metallicity and frequency of the Be phenomenon.  While metallicity clearly impacts single Be star formation, it also affects binary Be star formation because in low-metallicity environments, the progenitor stars retain more angular momentum and rotate faster, making it easier for binary interactions to spin up the star towards $v_{\rm crit}$.  

Metallicity may not be the only environmental factor that influences OB star rotation rates and decretion disk formation. In the Galaxy, OB stars rotate more slowly in the field than in clusters \citep[e.g.,][]{gut84, wol07, wol08}.  Field OB stars may rotate more slowly because they form in lower density environments characterized by lower turbulent velocities and lower infall rates than their cluster counterparts, resulting in the formation of stars with lower rotation rates \citep{wol07, wol08}.  Alternatively, field stars may be older, more evolved stars that have dispersed from their clusters and rotate more slowly because their radius and moment of inertia increase as they evolve \citep{hua08, hua10}.  Stellar winds also strip angular momentum from these stars, further slowing their rotation with age.  If the observed slower rotation in the field is an age effect, these arguments offer possible explanations, although we earlier cited arguments that rotation rates increase, rather than decrease, with age.  In any case, even in low-metallicity environments, if field OB stars are older, on average, the possibility exists that they may be less likely to rotate rapidly enough to form decretion disks.  In any case, given the higher frequency of rapid rotators in clusters, we might therefore expect the Oe/Be star frequency to be higher in clusters.  To date, few studies have observed how field and cluster environments affect decretion disk formation.  \citet{mar07b} measure the Be/B ratio in the SMC clusters to be 0.20--0.40 and $0.26\pm0.04$ in the field surrounding NGC~330, yielding inconclusive results.  

Although the frequency of Be stars has been studied in many environments, similar work on Oe stars is lacking, and few statistically complete spectroscopic samples of either Oe or Be stars exist.  Here, we use the Runaways and Isolated O Type Star Spectroscopic Survey of the SMC (RIOTS4), a spatially complete spectroscopic survey of SMC field OB stars \citep{lamin}, to identify a complete sample of classical Oe stars in the SMC field.  We obtain their spectral type distribution, measure their frequency, and evaluate the effects of metallicity and field environment.  In what follows, we present $29\pm1$ SMC field Oe stars, of which only two have previously been classified as Oe stars.  There are also two O6e stars, the earliest unambiguous SMC Oe stars identified to date.  Our observations demonstrate that Oe stars are the higher-mass analogs of the Be phenomenon, clarify the relationship between Ope stars and Oe stars, and reveal observational support for recent stellar evolution models of hot, rapidly rotating stars. 

\section{The RIOTS4 Survey}
\label{s:data}

RIOTS4 is a spatially complete, spectroscopic survey targeting 374 photometrically selected SMC field OB stars \citep{lamin}, identified from \citet{oey04}, with the goal of studying the nature of field OB stars.  To date, results include measurement of the field massive star IMF, yielding a power-law slope of $\Gamma=2.3\pm0.4$, where the Salpeter value is $\Gamma=1.35$ \citep{lam13}; evidence that some OB stars appear to form in the field \citep{lam10, oey13}; discovery of a class of dust-poor B[e] supergiants \citep{gra12}; and quantitative parameterization of the field OB population \citep{lamin}.

To identify SMC OB stars, RIOTS4 used two photometric criteria: $B\leq15.21$ and $Q_{UBR}\leq-0.84$, where $Q_{UBR}$ is the reddening-free parameter defined as \citep{oey04}:  
	\begin{equation}
	Q_{UBR}=(m_{U}-m_{R})-1.396(m_{B}-m_{R}) \quad .
	\end{equation}
In Equation 1, $m_{U}$, $m_{B}$, and $m_{R}$ are the apparent \textit{U}, \textit{B}, and \textit{R} magnitudes, respectively.  These criteria select stars with masses $\gtrsim 10~M_{\odot}$.  The \textit{B} magnitude criterion selects the most luminous stars, while the $Q_{UBR}$ parameter identifies the bluest stars, corresponding to spectral types of B0~V or B0.5~I and earlier \citep[e.g.,][]{oey04}.  As shown in Equation 1, $Q_{UBR}$ depends on the \textit{R}-band magnitude, which is sensitive to H$\alpha$.  The Balmer emission that characterizes Oe/Be stars therefore enhances the $Q_{UBR}$ criterion, promoting completeness of our Oe star sample. 

RIOTS4 identified field stars using a friends-of-friends algorithm \citep{bat91}, which defines field OB stars as stars located farther than a projected distance of 28 pc from any other OB candidate.  Thus, the RIOTS4 OB and Oe/Be stars represent a spatially complete sample of all field OB stars that meet the photometric selection criteria. 

The RIOTS4 spectroscopic observations were taken using the Magellan Telescopes at Las Campanas Observatory from 2006 to 2011 \citep[][]{lamin}.  Details about the observations and data reduction are described by \citet{lamin}, so here we present a summary.  The majority of spectra were taken using the IMACS spectrograph \citep{big03} in its multi-slit mode with a spectral resolution of \textit{R}~$\sim$~2600 or 3700.  A few additional observations were taken using the MIKE echelle spectrograph (\textit{R}~$\sim$28000) \citep{ber03}.  All RIOTS4 spectra cover the wavelength range 4000--4700~\AA; however, the majority extend to $\sim$5000~\AA, allowing us to identify Oe/Be stars using H$\beta$, in addition to H$\gamma$ and H$\delta$ emission. 

The IMACS multi-slit spectra were reduced using the COSMOS\footnote{\footnotesize{COSMOS was written by A. Oemler, K. Clardy, D. Kelson, G. Walsh, and E. Villanueva.  See http://code.obs.carnegiescience.edu/cosmos.}} data reduction software, and the 1D spectra were extracted using IRAF\footnote{\footnotesize{IRAF is distributed by the National Optical Astronomy Observatory, which is operated by the Association of Universities for Research in Astronomy (AURA), Inc. under cooperative agreement with the National Science Foundation (NSF)}}.  The MIKE and IMACS long-slit spectra were reduced and extracted using IRAF \citep{lamin}.  All reductions include background sky subtraction, and none of our Oe star spectra show evidence of any significant contamination from nebular line emission.  We also rectified each spectrum and removed cosmic rays.  In addition, for 13 low signal-to-noise Oe star spectra, we used the IRAF task {\tt bwfilter} to apply a Fourier filter and remove high-frequency noise.  We selected a filter frequency that preserves the double-peaked Balmer emission lines. 

\section{Spectral Classification}
\label{s:SpT}

We determine the spectral classifications of RIOTS4 O and Oe stars using the photospheric criteria of Walborn and coworkers.  For mid O-type stars, the principal diagnostic is the He~\textsc{ii}~$\lambda$4542/~He~\textsc{i}~$\lambda$4471 ratio, which is $\sim$1 for O7 stars \citep[e.g.,][]{wal90}.  For late O-types, we use the He~\textsc{ii}~$\lambda$4200/~He~\textsc{i}~$\lambda$4144 and He~\textsc{ii}~$\lambda$4542/~He~\textsc{i}~$\lambda$4387 ratios, which are $\sim$1 for O9 stars \citep[e.g.,][]{sot11}.  The former also distinguishes O9--9.5 stars from B0 stars, since the equivalent width of He~\textsc{ii}~$\lambda$4200 decreases for later-type O stars and is $\sim0$ for B0 stars \citep{wal90}.  In addition, the diagnostics for later O stars act as secondary indicators for early O and Oe stars because the He~\textsc{i}~$\lambda$4144 and He~\textsc{i}~$\lambda$4387 absorption lines do not appear in stars with spectral types earlier than O6--6.5 \citep[e.g.,][]{wal90}.  Si~\textsc{iv}~$\lambda\lambda$4089, 4116, Si~\textsc{iii}~$\lambda$4552, and C~\textsc{iii}~$\lambda$4650 are also standard diagnostics for late O and early-type B stars; however, these lines are often absent due to the low SMC metallicity.  We use them when available, but our classifications necessarily rely on He~{\sc i} and He~{\sc ii} lines. 

The lack of metal lines also causes problems in determining luminosity classes. The principal diagnostics are N~\textsc{iii}~$\lambda\lambda$4634-4640-4642, He~\textsc{ii}~$\lambda$4686, and the Si~\textsc{iv}~$\lambda$4089/~He~\textsc{i}~$\lambda$4026 ratio \citep{wal90}.  Thus, we also use extinction-corrected \textit{V} magnitudes from \citet{mas02} as an additional criterion, as well as line widths, especially for supergiants. 

We also include spectral type modifiers to further characterize our Oe stars.  Following the Be star classification system from \citet{les68}, we specify Oe stars whose spectra contain Fe~\textsc{ii} emission as O$\rm e_{\rm +}$ stars and we use the full \citet{les68} classification system, $\rm e_{\rm 1}$ to $\rm e_{\rm 4}$, based on Balmer emission line strength.  Following \citet{sot11}, we classify stars whose spectra show He~\textsc{i} infill or emission above the continuum as Ope stars, although we follow \citet{neg04} in using non-infilled \hei\ diagnostics to determine the spectral types of Ope stars.  For stars where the equivalent width of He~\textsc{ii}~$\lambda$4686 is clearly greater than the equivalent width of He~\textsc{i}~$\lambda$4471 and He~\textsc{ii}~$\lambda$4542, we add the \textit{z} qualifier \citep{wal97}.  Following the usual convention, ``:" connotes substantial uncertainty in the classification. 

\begin{figure}
\epsscale{1.2}
\plotone{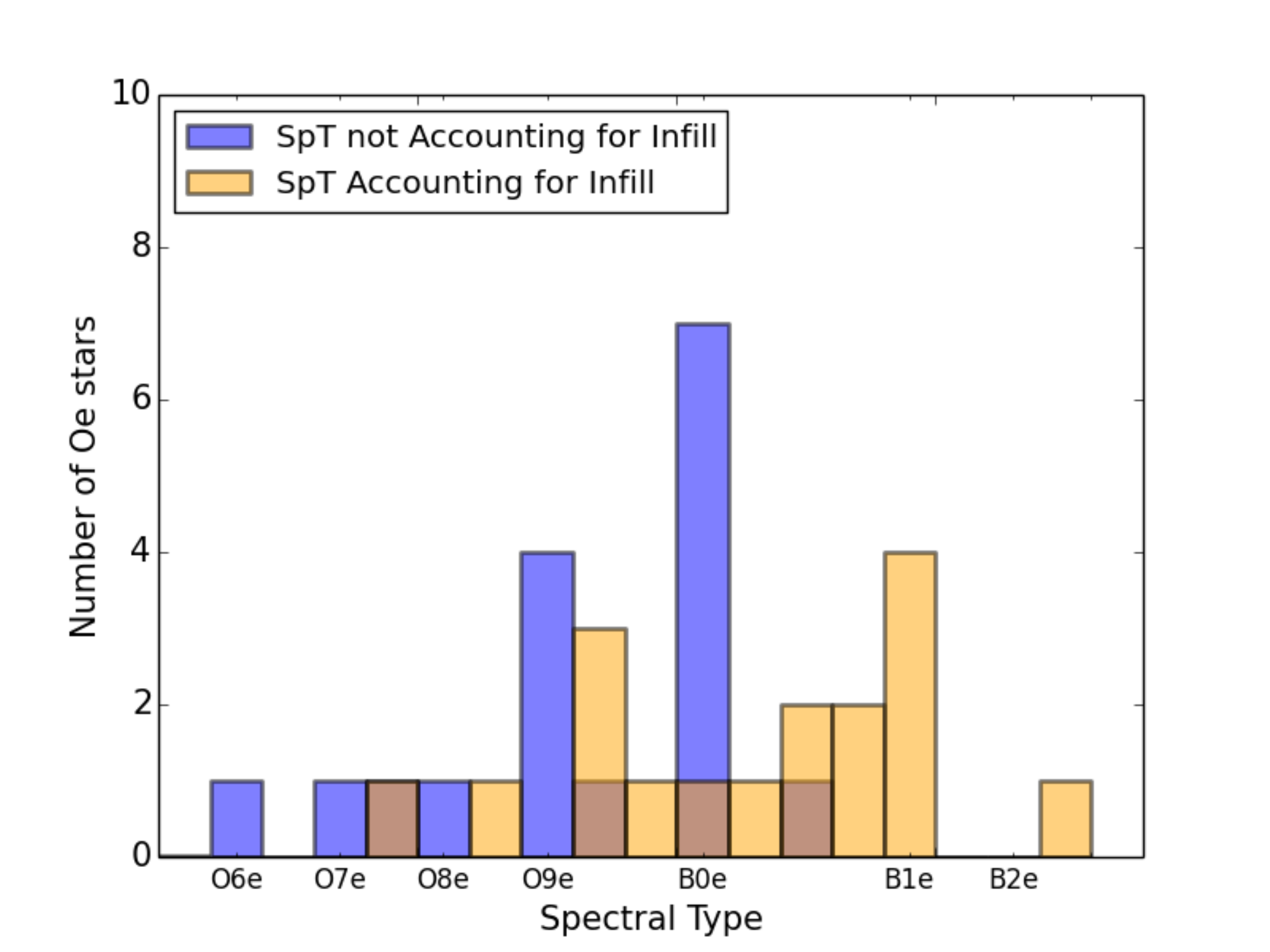}
\caption{Distributions of the spectral types of Oe/Be stars from \citet{neg04}.  The blue distribution represents the spectral types from the literature, which are based on the He~\textsc{ii}~$\lambda$4542/~He~\textsc{i}~$\lambda$4471 ratio, while the orange distribution represents the \citet{neg04} classifications obtained without using He~\textsc{i}~$\lambda$4471.}  
\label{f:Neg}
\end{figure}

The primary challenge when classifying our Oe stars comes from infill in the He~\textsc{i}~$\lambda$4471 absorption line by disk emission \citep{ste99}.  \citet{neg04} find that infill of \hei\ implies that many Oe stars are cooler than their prior classifications suggest.  Thus, they reclassified the majority of their Galactic Oe stars as O9.5e or later.  Figure~\ref{f:Neg} highlights this effect by comparing the spectral types presented in \citet{neg04}, determined without using He~\textsc{i}~$\lambda$4471, to prior classifications from literature of the same stars based on the He~\textsc{ii}~$\lambda$4542/~He~\textsc{i}~$\lambda$4471 ratio.  Given the importance of infill, we follow \citet{neg04} and classify our Ope stars without using criteria based on He~\textsc{i}~$\lambda$4471.  Since we cannot rely on metal-line diagnostics, we rely on the He~\textsc{ii}~$\lambda$4200/~He~\textsc{i}~$\lambda$4144 and He~\textsc{ii}~$\lambda$4542/~He~\textsc{i}~$\lambda$4387 ratios to determine their spectral types.  In such cases, we add a ``:" to our classification to connote that different classification criteria lead to different spectral types.    

At least three of the authors (JBGM, MSO, \& JBL) independently classified each RIOTS4 O and Oe star.  We compared our results, along with prior ones from literature, to converge on our final classifications for each star, which are accurate to within half a spectral type.  About 75\% of our classifications agree with values from the literature when available.  Table~\ref{t:alldat} presents the ID's from \citet{mas02} in column 1, and our spectral types in column 2.  Columns 3 and 4 provide existing spectral types from the literature and their sources, respectively.  The $V$-band magnitude \citep{mas02} is given in column 5, extinction $A_{V}$ \citep{zar02} in column 6, $J$-band magnitude \citep{skr06} in column 7, and $3.6\mu$m magnitude \citep{gor11} in column 8.  Our spectral types for all normal O stars are presented by \citet{lamin}.  In total, we identify 28--30 Oe stars in RIOTS4; the spectra are presented in the Appendix.  Only those with \citet{mas02} ID numbers 51373 \citep{mas95}, 72535 \citep{azz75}, and 75689 \citep{eva04} have been previously identified as Oe stars. 

Star 72535 was classified as O9~Ie by \citet{azz75}, and we obtain a spectral type of O8--9~I--IIIe.  This star has a high luminosity and is brighter in the $V$ and $J$-bands than any other RIOTS4 Oe star
(Table~\ref{t:alldat}).  As noted, supergiants are not considered classical Oe/Be stars.  Star 72535 also shows a high degree of variability for an Oe/Be star \citep[e.g.,][]{dew06,riv13}, at least 0.4 magnitudes in the 3.6, 4.5, and 5.4 $\mu$m bands over only a three month period in 2008, and 0.8 mag in the period from 2005 to 2008 \citep{gor11}.  The 2MASS $J$-band mag was observed in 1998 \citep{skr06} and not concurrently with the SAGE data, so colors cannot be inferred from these data.  Thus, this star does not appear to be a classical Oe star, and we do not include it in our Oe sample in what follows.

\begin{deluxetable*}{cccccccc}
	\tablecolumns{7}
	\tablewidth{0pt}
	\tabletypesize{\small}
	\tablecaption{RIOTS4 O\lowercase{$\rm e$} Stars}
	\tablehead{\colhead{ID\tablenotemark{a} }&
	\colhead{SpT\tablenotemark{b} }&
	\colhead{SpT}&
	\colhead{SpT Source\tablenotemark{c} }&
	\colhead{\textit{V}\tablenotemark{a} }&
	\colhead{$A_{V}$\tablenotemark{d} }&
	\colhead{\textit{J}\tablenotemark{e} }&
	\colhead{3.6$\mu$m\tablenotemark{f}} \\
	\colhead{}&
	\colhead{(this paper)}&
	\colhead{(literature)}&
	\colhead{}&
	\colhead{(mag) }&
	\colhead{(mag) }&
	\colhead{(mag) }&
	\colhead{(mag)} 
        } 
\startdata
7254 & O9.5 III$\rm e_{\rm 2}$ & \nodata & \nodata & 14.74 & 0.86 & 14.565 & 13.937 \\ 
11677 & O9 III:$\rm e_{\rm 3+}$ & \nodata & \nodata & 14.46 & 1.04 & 14.421 & 13.655 \\ 
12102 & O9 III$\rm e_{\rm 2}$ & O9.5 V & E04 & 14.69 & 0.84 & 15.113 & 14.332 \\ 
14324 & O6 V((f))$\rm e_{\rm 2}$ & \nodata & \nodata &14.11 & 1.12 & 13.719 & 13.031 \\ 
15271 & O6 III((f))$\rm e_{\rm 1}$ & O6 III((f)) & E04 & 13.54 & 0.39 & 13.999 & 14.105 \\ 
18329 & O9.5 III$\rm e_{\rm 4+}$ pec & \nodata & \nodata & 14.65 & 1.26 & 14.454 & 13.630 \\ 
22321 & O9.5 IIIp$\rm e_{\rm 4+}$ & \nodata & \nodata& 13.69 & 1.00 & 13.498 & 12.544 \\ 
23710 & O9--B0 p$\rm e_{\rm 3+}$ & \nodata & \nodata & 14.74 & 1.31& 14.334 & 13.503 \\ 
24914 & O9 III-Ve$_1$ & O2((f))+OBe & A09 & 14.19 & 0.95 & 13.682 & 13.011 \\
27884 & O7--8.5 Vp$\rm e_{\rm 4+}$ & \nodata & \nodata & 14.35 & 1.05 & 14.275 & 13.588 \\ 
30744 & O9.5 V$\rm e_{\rm 2}$ + B1 III & B0.5 IVe & E04 & 14.30 & 0.66 & 14.563 & 14.140 \\ 
37502 & O9.5--B0: p$\rm e_{\rm 3+}$ & \nodata & \nodata & 14.62 & 1.10 & 14.512 & 13.723 \\ 
38036 & O6.5--7: Vp$\rm e_{\rm 3}$ & B0 III & M07 & 14.33 & 0.82 & 14.342 & 13.884 \\ 
50095 & O9 V$\rm e_{\rm 4+}$ & \nodata & \nodata & 14.51 & 0.87 & 14.440 & 13.618 \\ 
51373 & O8 IIIz$\rm e_{\rm 3}$ & O7 Ve  & M95 & 13.79 & 0.84 & 13.685 & 13.531 \\ 
52363 & O7 III((f))$\rm e_{\rm 1}$ & O7 IIf & E04 & 14.98 & 0.38 & 15.576 & 15.771 \\ 
52410 & O8 III: z$\rm e_{\rm 3}$ pec & O8 V & M95 & 13.75 & 0.85 & 13.978 & 13.298 \\ 
53360 & O7--9 Vp:$\rm e_{\rm 2}$ & O6--9 & E04 & 15.09 & 0.77 & 15.187 & 14.427 \\ 
56503 & O9 V$\rm e_{\rm 2}$ & O9 III-V & E04 & 14.95 & 0.68 & 15.363 & 14.432 \\ 
59319 & early Op$\rm e_{\rm 3}$ & \nodata & \nodata & 14.32 & 0.68 & 14.322 & 13.435 \\ 
62638 & O9.5 III--V$\rm e_{\rm 2}$ & B0 IV & E04 & 14.67 & 0.65 & 14.696 & 14.562 \\ 
65318 & O9 V$\rm e_{\rm 1}$ & \nodata & \nodata & 14.95 & 0.35 & 15.147 & 14.918 \\ 
67673 & O9.5 V:$\rm e_{\rm 2}$ & \nodata & \nodata & 14.84 & 0.83 & \nodata & 14.974 \\ 
69460 & O6.5 III((f))$\rm e_{\rm 2}$ & O6.5 IIf & E04 & 14.95 & 0.53 & 15.121 & 14.345 \\ 
72535 & O8-9 I-IIIe\tablenotemark{g} & O9 Ie & A75 & 13.45 & 1.00 & 13.323 & 14.031 \\ 
73795 & mid O$\rm e_{\rm 3+}$ & O6--9 III-V & E04 & 14.73 & 0.82 & 14.492 & 13.650 \\ 
75689 & Ope$_3$\tablenotemark{h}  & O4-7 Ve & E04 & 14.34 & 0.94 & 14.174 & 13.319 \\ 
75919 & O9 III$\rm e_{\rm 1}$ & \nodata & \nodata & 14.39 & 0.78 & 14.441 & 14.209 \\ 
77616 & early Op$\rm e_{\rm 3}$ pec & B0 & A82 & 14.08 & 1.22 & 13.959 & 13.725 \\ 
78694 & O8.5 III$\rm e_{\rm 2+}$ & O9.5 III-V & E04 & 14.35 & 0.72 & 14.342 & 13.635 \\ 
82489 & O9: IIIp$\rm e_{\rm 4+}$ & \nodata & \nodata & 14.22 & 0.86 & 14.195 & 13.345
\enddata
\tablenotetext{a}{\footnotesize{\citet{mas02}}}
\tablenotetext{b}{\footnotesize{The subscripts in our spectral
    classifications are the \citet{les68} classifications of the Be phenomenon based on the
    strength of Balmer emission and the presence of Fe~\textsc{ii}
    emission.}} 
\tablenotetext{c}{\footnotesize{{A09: \citet{ant09}, A75:~\citet{azz75}, A82:~\citet{azz82}, E04:~\citet{eva04}, M07:~\citet{mar07}, M95:~\citet{mas95}}}}
\tablenotetext{d}{\footnotesize{\citet{zar02}}}
\tablenotetext{e}{\footnotesize{\citet{skr06}}}
\tablenotetext{f}{\footnotesize{\citet{gor11}}}
\tablenotetext{g}{\footnotesize{Not a classical Oe star (see text)}}
\tablenotetext{h}{\footnotesize{Probably a spectroscopic binary.}}
\label{t:alldat}
\end{deluxetable*}

\section{Results}   
\label{s:Results}

\subsection{The Earliest O\lowercase{e} and O\lowercase{pe} Stars}
\label{s:earlyOe}

\begin{figure*}
\epsscale{1.05}
\plotone{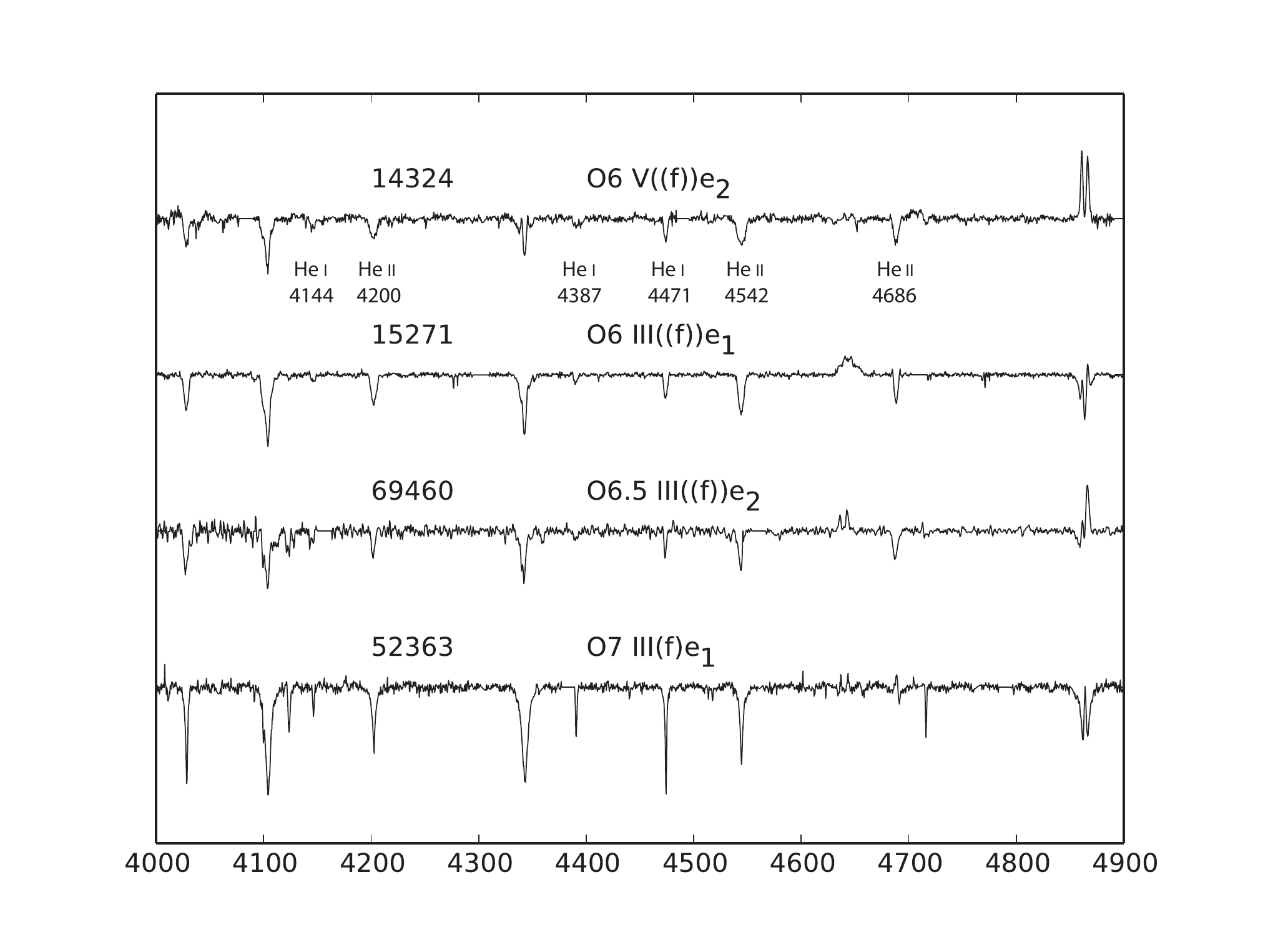}
\caption{The earliest classified RIOTS4 Oe stars.  The \citet{mas02} ID number and spectral type are listed above each spectrum.  The spectral sequence shows that the He~\textsc{ii}~$\lambda$4542/~He~\textsc{i}~$\lambda$4471 ratio decreases from earlier to later-type O stars, approaching $\sim1$ at O7 \citep{wal90}.  The absorption lines used in our classification are also labeled.} 
\label{f:earlyOe}
\end{figure*}

As described in \S~\ref{s:intro}, the decretion disk model predicts that in the low-metallicity SMC, earlier-type Oe stars can form than in the Galaxy, and RIOTS4 reveals some of the earliest known Oe stars to date.  Our four earliest have the classifications O6~V((f))$\rm e_{\rm 2}$, O6~III((f))$\rm e_{\rm 1}$, O6.5~III((f))$\rm e_{\rm 2}$, and O7~III(f)$\rm e_{\rm 1}$ (Figure~\ref{f:earlyOe}) and \citet{mas02} ID numbers 14324, 15271, 69460, and 52363, respectively.  These are the earliest unambiguous Oe star classifications reported in the SMC to date.  Of these, star 14324 appears to be a shell star, based on its Balmer absorption core.  As described in \S~\ref{s:SpT}, Oe stars are susceptible to \hei\ infill, especially \hei~$\lambda4471$ \citep{ste99}.  Thus, many mid and early-type Oe stars that exhibit infill are cooler than suggested by the spectral types obtained using the He~\textsc{ii}~$\lambda$4542/~He~\textsc{i}~$\lambda$4471 ratio (Figure~\ref{f:Neg}).  In the classification scheme of \citet{sot11}, such stars should be classified as Ope, but this has not been standardized, and infill can be overlooked.  Of particular interest, the MW Oe star HD~39680 has been classified as both O6 V:[n]pe \citep{sot11} and O8.5 Ve \citep{neg04}; the former corresponds to its spectral type based on the primary diagnostic affected by infill, while the latter is based on secondary diagnostics unaffected by \hei\ infill.  The earliest known Oe star in the Galaxy with no evidence of infill appears to be HD~155806, an O7.5~IIIe star \citep{neg04}.  To date, the earliest reported Oe star is Sk~--67~274 \citep{con86} in the LMC, classified as an O3e star.  No published spectrum of this star is available, so we are unable to evaluate whether this is an Ope star.  We stress that our four O6e -- O7e stars are certainly among the earliest known Oe stars that do \textit{not} show evidence of \hei\ infill.     

\begin{figure*}
\epsscale{1.05}
\plotone{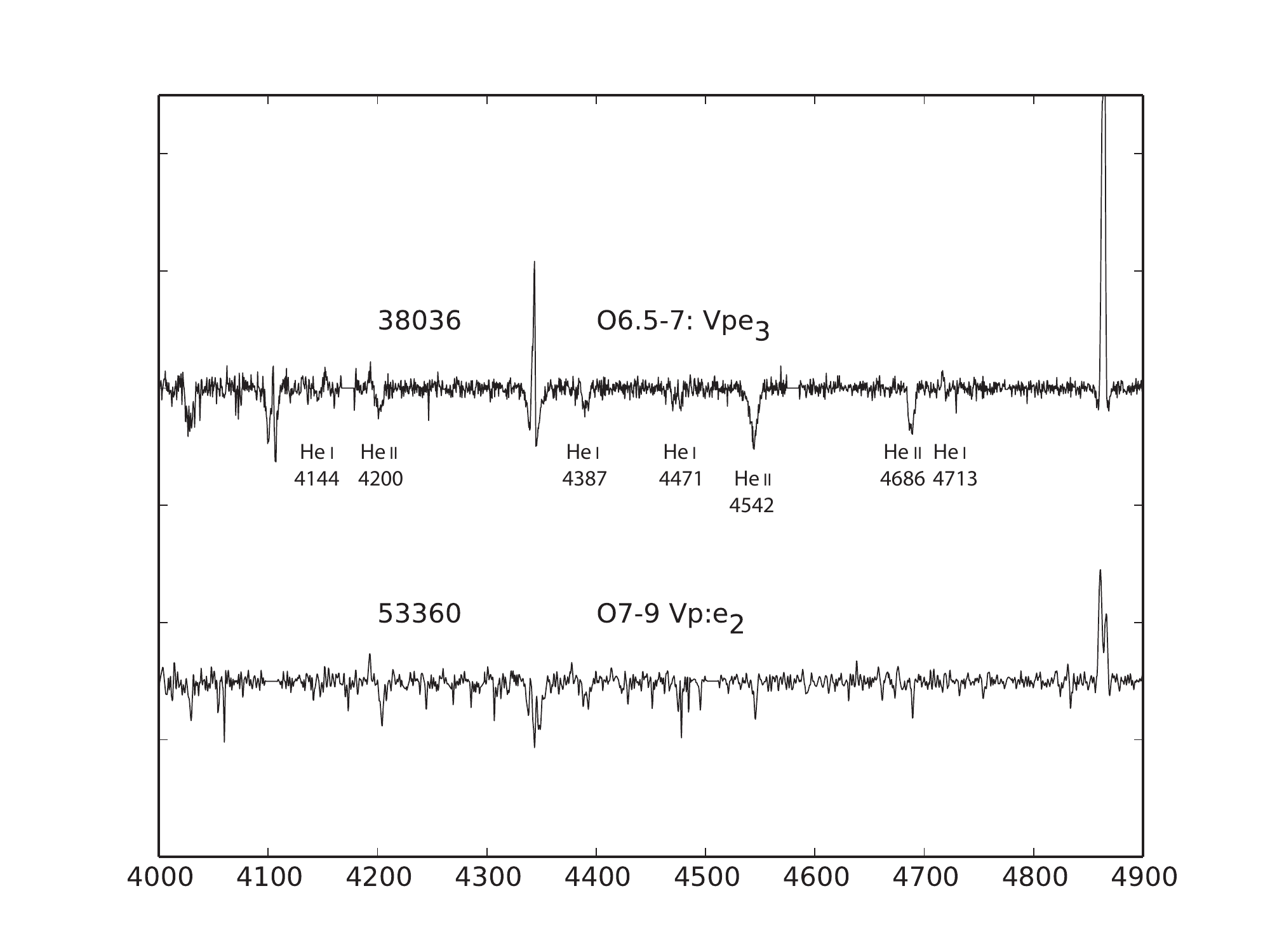}
\caption{Spectra of the RIOTS4 Ope stars that may be hotter than O7.5e stars; both show infill in He~\textsc{i}~$\lambda$4471.  The \citet{mas02} ID number and spectral type appear above each spectrum.  The absorption lines used to classify these stars and the He~\textsc{i} lines in infill/emission are also labeled.} 
\label{f:earlyOpe}
\end{figure*}  

In Figure~\ref{f:earlyOpe}, we present two Ope stars whose effective temperatures $\teff$ may be hotter than that of an O7.5e star.  Our spectral types are necessarily uncertain due to He~\textsc{i}~$\lambda$4471 infill, and we base our classifications on the He~\textsc{ii}~$\lambda$4200/~He~\textsc{i}~$\lambda$4144 and He~\textsc{ii}~$\lambda$4542/~He~\textsc{i}~$\lambda$4387 ratios.  The first star, 38036, contains He~\textsc{i}~$\lambda$4471 and $\lambda$4713 infill, so we use the He~\textsc{ii}~$\lambda$4200/~He~\textsc{i}~$\lambda$4144 and He~\textsc{ii}~$\lambda$4542/~He~\textsc{i}~$\lambda$4387 ratios to classify this star.  The absence of He~\textsc{i}~$\lambda$4144 and presence of He~\textsc{i}~$\lambda$4387 imply a spectral type of O6.5--7:~Vp$\rm e_{\rm 3}$.  The low signal-to-noise of the second star, 53360, may allow for infilled He~\textsc{i}~$\lambda$4471 and $\lambda$4387.  Thus, we use the He~\textsc{ii}~$\lambda$4200/~He~\textsc{i}~$\lambda$4144 ratio to estimate that star 53360 is an O7--9 Vp:$\rm e_{\rm 2}$ star.    

\begin{figure*}
\epsscale{1.05}
\plotone{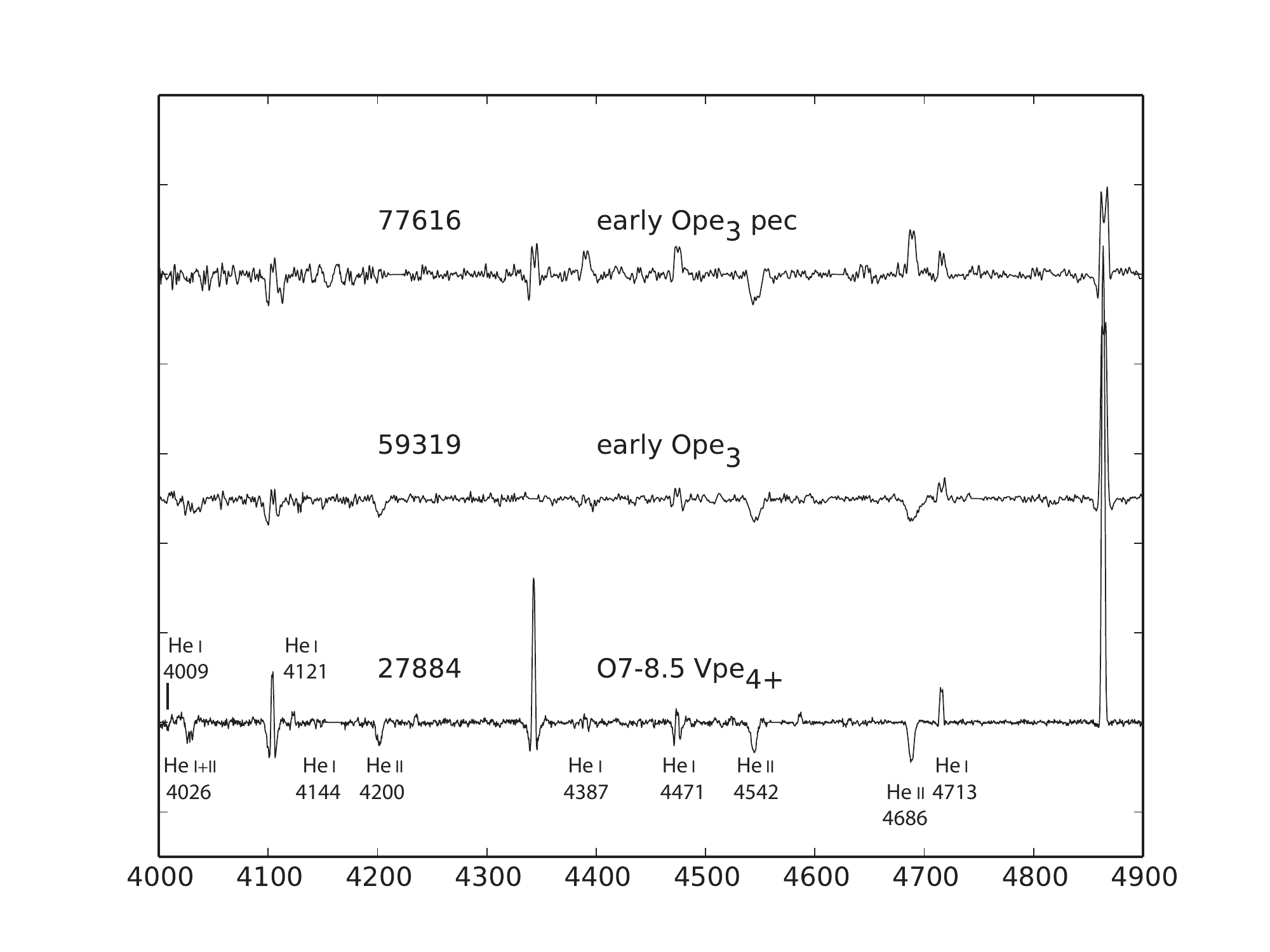}
\caption{The three extreme early-type RIOTS4 Ope stars, all with He~\textsc{i}~$\lambda$4713 in emission.  The \citet{mas02}~ID number and spectral type appear above each spectrum.  The He~\textsc{i} and He~\textsc{ii} features in infill and emission are also labeled.} 
\label{f:earlyOpec}
\end{figure*} 

Figure~\ref{f:earlyOpec} shows three additional early to mid-type Ope stars, which have stronger \hei\ emission than other Ope stars.  In particular, all have He~\textsc{i}~$\lambda$4713 and $\lambda$4471 in emission.  Furthermore, these stars show evidence of infill in the majority of \hei\ lines.  Star 59319 shows essentially no \hei\ absorption, and based on the equivalent widths of He~\textsc{ii}~$\lambda$4200 and $\lambda$4542, this is an early Op$\rm e_{\rm 3}$ star, likely earlier than O7.5e.  Star 27884 could be as late as O8.5:~Vp$\rm e_{\rm 4+}$, based on the He~\textsc{i+ii}~$\lambda$4026/~He~\textsc{i}~$\lambda$4009 ratio; the absence of C~\textsc{iii}~$\lambda$4650 sets our earliest estimate for the spectral type, O7:~Vp$\rm e_{\rm 4+}$.  In addition, we observe Fe~\textsc{ii}~$\lambda\lambda$~4233, 4583 clearly in emission.  The spectrum of 27884 represents the composite of ten observations obtained over four years for binary monitoring, and radial velocity variations suggest that this star may indeed be a binary \citep[see][]{lamin}.    

\subsubsection{Star 77616: He~\textsc{ii}~$\lambda$4686 Disk Emission}
The third star in Figure~\ref{f:earlyOpec}, 77616 \citep[AzV 493;]{azz82}, may be the most extreme early Oe/Ope star.  The spectrum shows strong \hei\ $\lambda4387$, $\lambda4471$, and $\lambda4713$ emission, and is remarkable because it is the only Oe star with He~\textsc{ii}~$\lambda$4686 emission from the Be phenomenon.  He~\textsc{ii}~$\lambda$4686 emission is associated with stellar winds in Of stars, but the double-peaked line profile here is clearly associated with the circumstellar disk.  Furthermore, we do not observe the characteristic N~\textsc{iii}~$\lambda\lambda$4634-40-42 emission of Of stars.  Double-peaked \heii\ emission is also associated with Onfp stars, which, as described in \S~\ref{s:intro}, have been proposed as high-mass analogs of the Be phenomenon \citep{con74}.  However, Onfp spectra show Balmer emission only in H$\alpha$, whereas the spectrum of star 77616 has strong \hei\ infill/emission.  Since star 77616 apparently represents the extreme Oe/Be phenomenon at the earliest types, and shows spectral properties that differ strongly from Onfp stars, this further confirms that Oe stars, and not Onfp stars, are the high-mass analogs of the Be phenomenon.  \heii\ $\lambda$4686 emission is also observed in the spectrum of MWC 656, the Be-black hole X-ray binary \citep{cas14}.  In MWC 656, He~\textsc{ii} emission is associated with the accretion disk surrounding the black hole.  No X-ray source is identified at the location of star 77616 in the XMM survey of the SMC \citep{hab12}, so it is unlikely that a similar origin can explain the \heii\ emission in 77616.  Additionally, there is no discrepancy between the radial velocity of the He~\textsc{ii}~$\lambda$~4686 emission and the other emission lines.

The He~\textsc{i} and He~\textsc{ii} infill/emission in Ope stars result from $\teff$ high enough to singly or doubly photoionize He in the disks.  In the case of star 77616, He~\textsc{ii} emission suggests an extremely high $\teff$.  Estimates for ionizing sources of nebular He~\textsc{ii}~$\lambda$4686 emission require $T_{\rm eff}\geq60,000$~K \citep[e.g.,][]{gar91}, which is extremely rare for main sequence stars.  However, models by \citet{bro11} suggest that some rapidly-rotating, chemically-homogeneous stars reach this temperature.  Oe stars are hot, rapidly rotating massive stars, so they offer an important test of the \citet{bro11} models.  These models predict that at the SMC metallicity, stars with masses $\ge 30~M_{\odot}$ and initial rotation velocities $>400$ km/s reach $T_{\rm eff}\sim 60,000$ K at some point in their evolution.  The existence of star 77616 is therefore an exciting discovery supporting these predictions.  Extremely hot O stars in low-metallicity environments have been proposed as ionizing sources for the He~\textsc{ii} emission in H~\textsc{ii} regions that is occasionally observed, especially in extreme starburst galaxies \citep{kud02}.  Thus, the earliest Oe/Ope stars, like star 77616, may be responsible for this emission. 

\subsubsection{Ope Stars: Normal Early-Type Oe Stars}
As previously discussed, the \hei\ and \heii\ emission that characterize Ope spectra require high temperatures; however, Table~\ref{t:Opetab} shows that strong disk emission is also necessary.  The distribution of \citet{les68} classifications and spectral types in Table~\ref{t:Opetab} shows that nine of ten Ope stars, including star 75689, are $\rm Oe_{\rm 3}$ or $\rm Oe_{\rm 4}$ stars, which have the strongest Balmer emission, and thus the strongest disk emission.  The only Ope$_1$ -- Ope$_2$ star has uncertain Ope status: star 53360 has ambiguous He~\textsc{i} infill and may not be an Ope star (Figure~\ref{f:earlyOpe}).  In total, fifteen RIOTS4 Oe stars, including the nine Ope stars, are Oe$_3$ or Oe$_4$ stars, and the six remaining Oe stars have spectral types of O7.5e or later. Furthermore, our earliest ordinary Oe stars have \citet{les68} classifications of $\rm Oe_{\rm 1}$ or $\rm Oe_{\rm 2}$, which supports the scenario that the photosphere is visible in early-type Oe stars only if the disk emission is weak.  This may be due to high angle of inclination, or physically smaller disks.  We do note a tendency for Oe$_1$ and Oe$_2$ stars to show double peaked Balmer emission (Figure~\ref{f:Appendix2}), indicative of high inclinations.  Table~\ref{t:Opetab} illustrates that among RIOTS4 $\rm Oe_{3}$ and $\rm Oe_{\rm 4}$ stars, only Ope stars are observed at early spectral types.  Therefore, depending on disk emission, it appears that classical Oe stars transition to Ope stars at early spectral types, in particular around type O7e and earlier. 

\begin{deluxetable*}{ccccccc}
	\tablecolumns{4}
	\tablewidth{0pt}
	\tabletypesize{\small}
	\tablecaption{Frequency of O\lowercase{$\rm pe$} stars among RIOTS4 O\lowercase{$\rm e$} stars\tablenotemark{a}}
	\tablehead{\colhead{ }&
	\colhead{early--O7e}&
	\colhead{O7.5e--8.5e}&
	\colhead{O9e--9.5e}}
\startdata
Oe$_{1, 2} $ & $0.0\pm0.0$ (0,4) & $0.50\pm0.35$ (1,2) & $0.11\pm0.10$ (1,9) \\
Oe$_{3, 4} $ & $1.0\pm0.0$ (3,3) & $0.25\pm0.22$ (1,4) & $0.57\pm0.19$ (4,7) \\
All Oe & $0.43\pm0.20$ (3,7) & $0.33\pm0.19$ (2,6) & $0.31\pm0.12$ (5,16) 
\enddata 
\tablenotetext{a}{The Ope/Oe ratio, where the denominator includes both Oe and Ope stars in each bin.  The numbers in parentheses correspond to the numbers of Ope and Oe~+~Ope stars, respectively.  Star 73795 is included in the O7.5e--8.5e category, and star 75689 is excluded from the table.}
\label{t:Opetab}
\end{deluxetable*}

\subsubsection{Fe~\textsc{ii} Emission in Oe Stars}
In addition to He~\textsc{i} emission, many Oe stars display Fe~\textsc{ii} emission lines in their spectra.  The most common emission lines are Fe~\textsc{ii}~$\lambda\lambda$4233, 4583, and 4629.  Similar to the trends with He~\textsc{i} emission shown in Table~\ref{t:Opetab}, Table~\ref{t:FeOetab} shows that Fe~\textsc{ii} emission also correlates with \citet{les68} classification and spectral type.  All but one Fe~\textsc{ii} emitting Oe star is a \citet{les68} Oe$_{\rm 3}$ or Oe$_{\rm 4}$ star.  We also find that Fe~\textsc{ii} emission is absent among Oe stars earlier than O7.5e.  Interestingly, the frequency of Fe~\textsc{ii} emission appears to stay constant at approximately 0.50 among all Oe stars later than O7.5e.  Using the preliminary classifications of RIOTS4 Be stars \citep{lamin}, we also find the same Fe~\textsc{ii} emission lines and the same frequency of Fe~\textsc{ii} emission.  Based on the trends in Tables~\ref{t:Opetab} and \ref{t:FeOetab}, the decrease in the $\teff$ of O stars promotes Fe~\textsc{ii} emission for stars strong emission. 

\begin{deluxetable*}{ccccccc}
	\tablecolumns{4}
	\tablewidth{0pt}
	\tabletypesize{\small}
	\tablecaption{Frequency of Fe~\textsc{ii} emission among RIOTS4 O\lowercase{$\rm pe$} stars\tablenotemark{a}}
	\tablehead{\colhead{ }&
	\colhead{early--O7e}&
	\colhead{O7.5e--8.5e}&
	\colhead{O9e--9.5e}}
\startdata
Oe$_{1, 2} $ & $0.0\pm0.0$ (0,4) & $0.50\pm0.35$ (1,2) & $0.0\pm0.0$ (0,9) \\
Oe$_{3, 4} $ & $0.0\pm0.0$ (0,3) & $0.50\pm0.25$ (2,4) & $1.0\pm0.0$ (7,7) \\
All Oe & $0.0\pm0.0$ (0,7) & $0.50\pm0.20$ (3,6) & $0.44\pm0.12$ (7,16)
\enddata 
\tablenotetext{a}{The frequency of Oe stars with Fe~\textsc{ii} emission, where the denominator includes the total number of Oe stars in each bin.  The numbers in parentheses are the number of Oe stars with Fe~\textsc{ii} emission and the total number of Oe stars.  As before, star 73795 is included in the O7.5e--8.5e category, and star 75689 is excluded from the table.}
\label{t:FeOetab}
\end{deluxetable*}

\subsection{Spectral Type Distribution of O\lowercase{$e$} Stars}
\label{s:Oedist}

\begin{figure}
\epsscale{1.2}
\plotone{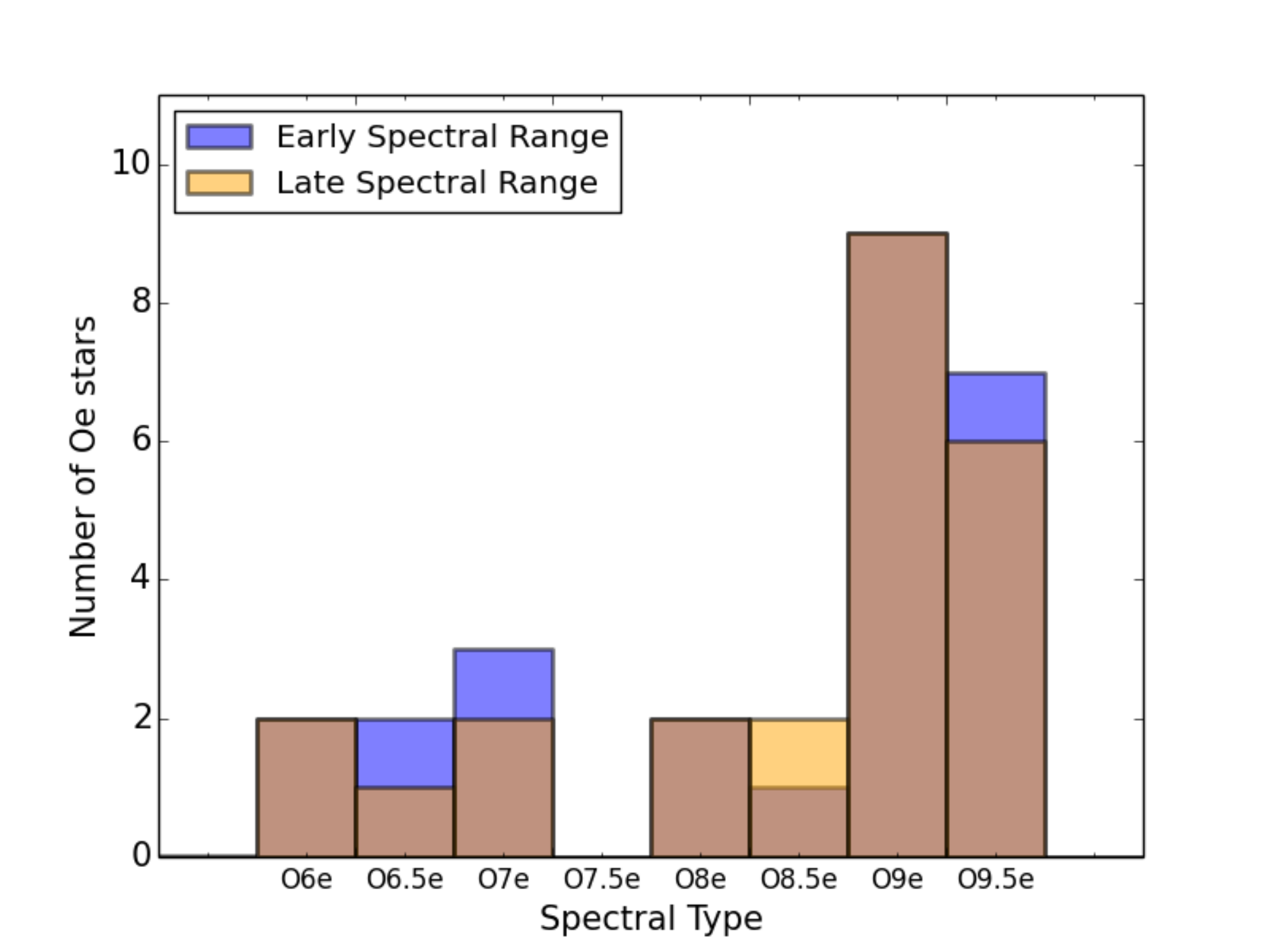}
\caption{The number of Oe stars observed at each spectral type.  To account for uncertainty in the spectral types of $\sim$20\% of our stars, the two distributions represent those obtained using our earliest estimate (blue) and latest (orange) for stars with uncertain spectral types.  Stars without spectral types are not included in this figure.}  
\label{f:dist}
\end{figure}  

We present the spectral type distribution of RIOTS4 Oe stars in Figure~\ref{f:dist}.  Since $\sim$20\% of our Oe stars have uncertain classifications, we present two distributions, based on the earliest (30 stars; blue) and latest (28 stars; orange) possible spectral type estimates for the uncertain classifications.  The number of stars in these distributions differ because the latest possible spectral type for some stars fall into the Be range, which is not included in Figure~\ref{f:dist}.  Additionally, Figure~\ref{f:dist} does not include the two Ope stars and two Oe stars that we are unable to classify.

Figure~\ref{f:dist} shows that 53--54\% of RIOTS4 Oe stars have spectral types of O9--9.5e.  Additionally, stars earlier than O7.5e, including the early Ope stars, account for 21--27\% of our sample, unlike in the MW where, to date, no Oe stars in this range have been observed \citep[e.g.,][]{neg04, sot11, sot14, sot14b}.  For the same spectral type distribution as the SMC, we would expect that $2\pm1$ of the eight MW Oe stars should have spectral types earlier than O7.5e.  Thus, our results demonstrate that, compared to the MW, the distribution of Oe stars in the SMC field is significantly enhanced towards earlier-type stars.  We caution that \citet{mcb08} find that the spectral type distributions of Be/X-ray binaries in the SMC and MW are similar, whereas \citet{neg98} finds quite different distributions between Be stars and Be/X-ray binaries in the MW.  

We calculate the Oe/O ratio using the 28--30 Oe stars we identify and the 106--109 RIOTS4 O~stars \citep{lamin}, which include Oe stars and exclude supergiants.  As noted, ambiguity in the spectral types of stars whose classifications allow them to fall into the early B-star range causes the uncertainty.  Table~\ref{t:freq} shows the Oe/O ratios and binomial errors at each spectral type.  Columns~2 and 3 show the values based on the earliest and latest possible spectral types for stars with uncertain classifications.  We find that the frequency of Oe stars is approximately constant between $\sim$0.15 -- 0.30 with a possible increase at the earliest spectral types, although we caution that stochastic effects are worst for the earliest bin. 

\begin{deluxetable}{ccc}
	\tablecolumns{3}
	\tablewidth{0pt}
	\tabletypesize{\small}
	\tablecaption{O\lowercase{$\rm e$}/O Ratio at each Spectral Type\tablenotemark{a}}
	\tablehead{ \colhead{SpT} &
	\colhead{Oe/O} &
	\colhead{Oe/O} \\ 
	\colhead{} & 
	\colhead{Earliest} &
	\colhead{Latest}}
\startdata
O6--6.5 & $0.57\pm0.19 $ (4) & $0.50\pm0.20$ (3) \\ 
O7--7.5 & $0.21\pm0.11$ (3) & $0.17\pm0.11$ (2) \\ 
O8--8.5 & $0.11\pm0.06 $ (3) & $0.14\pm0.07$ (4) \\ 
O9--9.5 & $0.30\pm0.06 $ (16) & $0.29\pm0.06$ (15)
\enddata
\tablenotetext{a}{The number of Oe stars in each spectral type range is listed in parentheses.  The stars we are unable to classify are not included.} 
\label{t:freq}
\end{deluxetable}
      
\subsection{Metallicity and the Frequency of Oe stars}
\label{s:met}

We measure an Oe/O ratio of $0.26\pm0.04$ ($0.27\pm0.04$) when considering the latest (earliest) possible spectral types for RIOTS4.  We adopt the value from the latest spectral types, in what follows.  We compare our Oe star frequency to an estimate by \citet{bon10} based on photometric identifications of a sample of 208 previously classified O stars, in which Oe star candidates are identified using IR photometry from the the IR Survey Facility (IRSF) Magellanic Clouds Point Source Catalog \citep{kat07} and \textit{Spitzer} SAGE-SMC survey \citep{gor11}.  Although the stars have existing spectral types, \citet{bon10} use NIR photometry to more consistently identify Oe star candidates in their heterogeneous sample.  They obtain an Oe/O ratio of $0.10\pm0.02$ using the selection criteria $J_{\rm IRSF}-[3.6]>0.5$ and $J_{\rm IRSF}<15$.  

We find that 12 of the 16 Oe star candidates \citet{bon10} identified were spectroscopically observed in H$\alpha$ and show emission \citep{eva04, mey93}, although one is a supergiant, which cannot be a classical Oe/Be star.  We confirm that two H$\alpha$-emitters are RIOTS4 Oe stars and one we classify as a Be star.  Of the four remaining candidates, one is a RIOTS4 Oe star, two show no Balmer emission in existing spectra from the literature, and one shows only nebular emission \citep{eva04}.  Thus, 75\% of the \citet{bon10} candidates are likely Oe stars, a reasonably good return. 

\begin{figure}
\epsscale{1.2}
\plotone{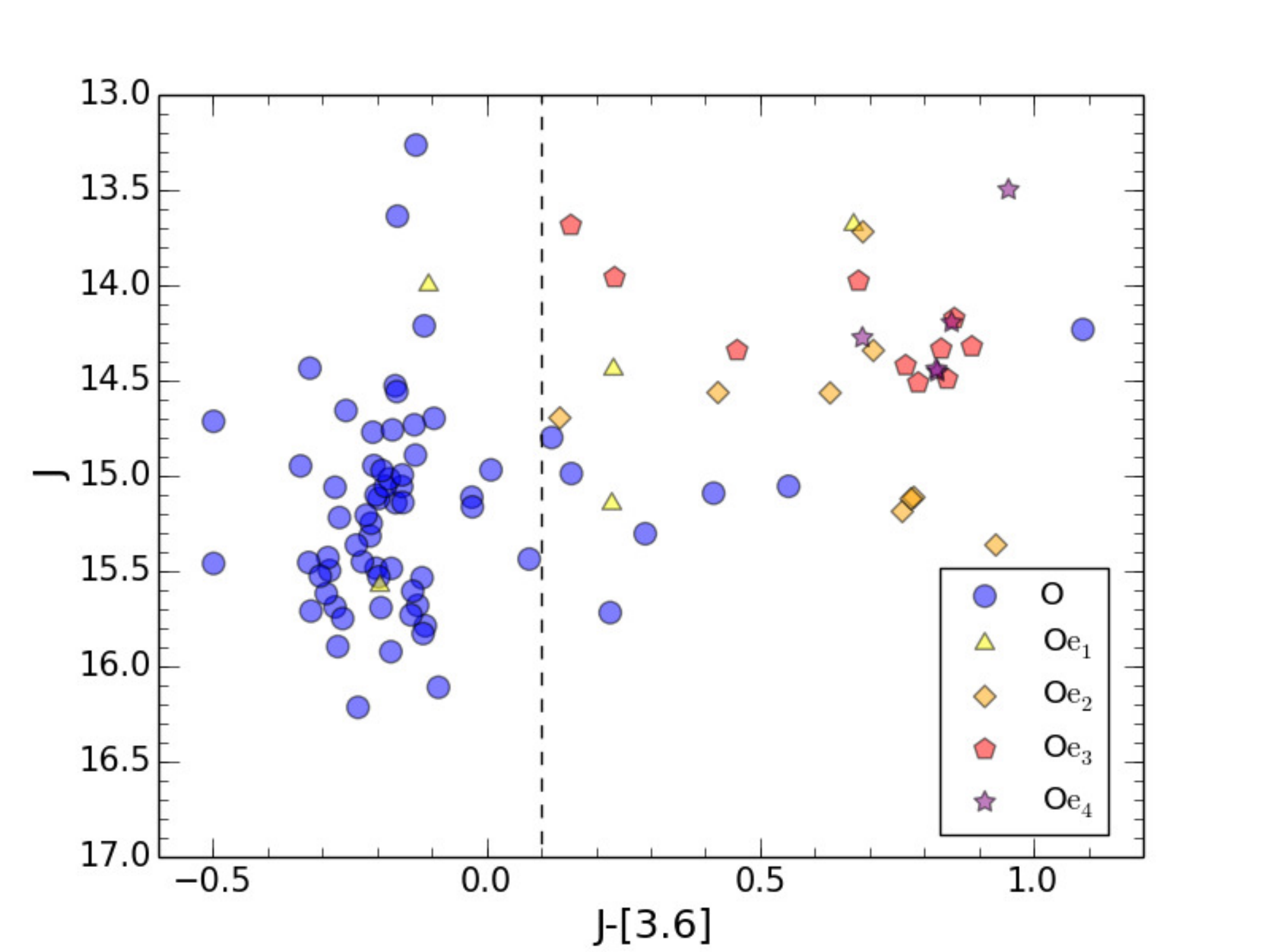}
\caption{IR color-magnitude diagram for RIOTS4 O and Oe stars.  The key shows symbols for normal O stars and Oe stars, divided by their \cite{les68} classifications.  The vertical dashed line shows our revised Oe star candidate photometric criterion, $J-[3.6]\geq0.1$.  We caution that the $J$ and [3.6] data were observed many years apart, so the colors have significant uncertainty due to possible variability (see text).} 
\label{f:CMD}
\end{figure} 

However, the \citet{bon10} color criterion, based on the $J-[3.6]$ colors of four previously identified SMC Oe stars, causes the discrepancy between their Oe/O ratio and ours.  Figure~\ref{f:CMD} shows the $J$ vs $J-[3.6]$ color-magnitude diagram for RIOTS4 O and Oe stars, using $J$-band magnitudes from the Two-Micron All Sky Survey \citep[2MASS;][]{skr06}, and $3.6\mu$m-band magnitudes from the \textit{Spitzer} SAGE-SMC survey \citep{gor11} (Table~\ref{t:alldat}).  One Oe and ten O stars are omitted from Figure~\ref{f:CMD} because the IR magnitudes are unavailable.  Figure~\ref{f:CMD} shows that the \citet{bon10} criteria exclude 13 of our 28~$\pm 1$ spectroscopically identified Oe stars with IR data.  We therefore propose a revised criterion, $J-[3.6]\geq0.1$, to select Oe candidates, which accounts for all but two RIOTS4 $\rm Oe_{\rm 1}$ stars, both of which have infilled Balmer lines, corresponding to the weakest presentation of the Oe/Be phenomenon \citep{les68}. 

Applying our $J-[3.6]$ criterion to the \citet{bon10} sample increases the number of their Oe star candidates from 16 to 34, and raises their Oe star frequency to $0.22\pm0.03$.  Similarly, applying our color criterion to RIOTS4, we identify 7 new RIOTS4 Oe star candidates.  These may or may not be Oe stars; their RIOTS4 spectra show no H$\beta$, H$\gamma$ or H$\delta$ emission, but they may have H$\alpha$ emission, which is beyond our observed spectral range.  Additionally, the strength of the Be phenomenon may have varied between when the \textit{Spitzer} SAGE-SMC and RIOTS4 observations were taken.  In any case, including all 7 candidates raises our Oe/O ratio to $0.35\pm0.05$ from our spectroscopic value, $0.26\pm0.04$.  We note that our formal uncertainties still show agreement between our photometric selection and spectroscopic identifications.  There are a few stars in Figure~\ref{f:CMD} whose classifications are inconsistent with the colors according to our criterion, and these may have incorrect $J-[3.6]$ colors if the magnitudes varied between the two photometric observations.  The 2MASS $J$-band observations were obtained between 1997--2001 \citep{skr06}, while the the SAGE 3.6$\mu$m data were observed in 2005 and 2008 \citep{gor11}. We note that our classical Oe stars do not show significant variability in their $3.6 \mu$m magnitudes.  The modest discrepancy between our photometrically determined frequency and the revised \citet{bon10} value may be caused by such color errors and differences in sample selection.  RIOTS4 is a complete sample of SMC field O stars, while the \citet{bon10} sample is heterogeneous, selecting all O stars with existing spectral classifications. 

Figure~\ref{f:CMD} shows that O stars have $J-[3.6]\sim -0.2$, as expected from the Rayleigh-Jeans portion of an O star blackbody curve, whereas Oe stars are clustered near $J-[3.6]\sim 0.8$.  Additionally, the $J-[3.6]$ colors of our Oe stars match those of spectroscopically confirmed SMC Be stars from Figure 10 of \citet{bon10}.  The only difference between Oe and Be stars in these bands is that Be stars have $J$ magnitudes between 14 and 18, while Oe stars have $J\sim$~13 to 15, highlighting that Oe stars are the more luminous extension of the Be phenomenon.  In Figure~\ref{f:CMD}, it is especially interesting that stars concentrated in the clump at $J-[3.6]\sim 0.8$ have the highest-value \citet{les68} classifications, $\rm Oe_{\rm 3}$ and $\rm Oe_{\rm 4}$, corresponding to the strongest Be phenomena.  Thus, these stars must have the largest disks.  The range in $J-[3.6]$ values presumably results from disks of different sizes, which may correspond to different growth stages.  Disk viewing angle also plays an important role.

Figure~\ref{f:CMD} also shows that Oe stars are on average approximately 1 magnitude brighter than O stars in the $J$-band, where the flux from the circumstellar disk strongly dominates over that from the stellar photosphere.  The excess flux has been observed to reach up to half a magnitude in the visible bands and dominates in the IR \citep[e.g.,][]{riv13}.  We created a simple model of the combined blackbody emission of an SMC O9.5 star with $\teff \sim$ 31,000 K and log~$g \sim4.2$ \citep{geo13} and a disk with emission approximated using the thin disk model of \citet{ada87}.  Using this model, we reproduce a 1 mag $J$-band excess and a 0.2 magnitude $V$-band excess when adopting inner and outer radii of 7$R_*$ and 10$R_*$, respectively, where $R_*$ is the stellar radius.  This is not meant to be a realistic representation of a Be-star disk, but serves to illustrate that the observed thermal emission properties are consistent with reasonable parameters.  Additionally, the somewhat evolved nature of Oe/Be stars may enhance their total luminosities.  

Using the \citet{bon10} sample of OB stars, \citet{kou14} also attempt to identify Oe candidates, using photometric light curves from OGLE~III \citep{uda08}, a $VI$ photometric survey of the SMC.  Their criteria are based on irregular variability and a color criterion, yielding an Oe/O ratio of $0.13\pm0.02$.  Since they use the same sample as \citet{bon10}, the color criterion suggests that a similar revision may rescale the \citet{kou14} frequency to a value more consistent with our own.

In the Galaxy, only eight Oe stars are found in the GOSSS survey \citep{sot11, sot14, sot14b}, yielding a frequency of spectroscopically identified Oe stars of $0.03\pm0.01$, far less than our spectroscopic SMC field measurement, $0.26\pm0.04$.  We caution that the MW estimate may be missing some Oe stars, as mentioned in \S~\ref{s:intro}, and that comparison of the GOSSS and RIOTS4 statistics is problematic due to the different sample selection; GOSSS is a Galactic, magnitude-limited survey, while RIOTS4 is a spatially complete sample of SMC field OB stars.  Given the lack of Oe star samples, the GOSSS estimate is one of the few we can compare to our measurements.  Keeping in mind the possible biases, the relative incidence of Oe stars in these two environments strongly supports the decretion disk model.  Our results are also similar to the trend reported by \citet{mar10}, that the Be phenomenon is enhanced by a factor of 3--5 among earlier-type OB stars in the SMC, compared to the MW. 

As mentioned in \S~\ref{s:earlyOe}, the LMC O3e star, Sk~--67~274 \citep{con86}, may be the earliest-type Oe star currently known.  While we did not identify an O3e star in the SMC, this galaxy has fewer stars than the LMC, so the probability of hosting an O3e star is lower.  We evaluate the likelihood of observing an O3e star in the LMC by comparing the frequency of Oe and Be stars in the MW, LMC, and SMC.  The Be/B ratios are 0.17$\pm 0.03$ \citep[][see \S~\ref{s:intro}]{zor97}, $0.175\pm0.025$ \citep{mar06}, and $0.26\pm0.04$ \citep{mar07b} in the fields of these galaxies, respectively.  Similarly, we find that the Oe/O ratios are $0.03\pm0.01$ in the MW and $0.26\pm0.04$ in the SMC field.  We use the ratios between the Be and Oe star frequencies and the trend with metallicity to estimate a lower limit on the frequency of Oe stars in the LMC of $\sim0.06$.  There are approximately two dozen O3 stars in the LMC, excluding the 30 Dor region \citep{ski14, wal14}.  If we assume the frequency of Oe stars is constant across all spectral types, then we expect to find one O3e star in the LMC.  Thus, the observed O3e star in the LMC is consistent with the metallicity trend, without considering the status of Ope stars.

\subsection{Field vs. Cluster Environment and the Frequency of O\MakeLowercase{e} Stars}
\label{s:fvc}
Although we find that our SMC field Oe/O ratio is much greater than the estimate for the MW, we again caution that our sample contains only field stars, which may bias a comparison to the GOSSS sample.  As described in \S~\ref{s:intro}, on average, Galactic cluster OB stars rotate faster than their counterparts in the field \citep{gut84, wol07, wol08}, which should cause the frequency of Oe/Be stars to be higher in clusters.  Turning to the SMC, \citet{mar10} measure the frequency of Oe stars in SMC clusters to be $0.24\pm0.09$\footnote{\footnotesize{\citet{mar10} do not present an uncertainty for this value, so we calculate the binomial error from the data in their paper.}} among O8--9 stars, binned in integer spectral types, which agrees well with our Oe/O ratio of $0.24\pm0.05$ for the same spectral range.  We caution that the spectral types from \citet{mar10} are rough estimates based on $BVI$ colors and the absolute $V$ magnitude.  However, the similarity between these frequencies is consistent with measurements for the Be/B ratio in the SMC field and clusters from \citet{mar07b}: for SMC clusters and the field surrounding NGC~330, the Be/B ratios are estimated to be 0.20--0.40 and $0.26\pm0.04$ respectively.  When examining the average rotational velocities for field and cluster B~stars, \citet{mar07} find no significant difference, obtaining $159\pm20$ km/s and $163\pm18$ km/s, respectively.  Thus, unlike in the Galaxy, in the SMC, B stars do not rotate more slowly in the field than in clusters, which is consistent with the similarity in the frequency of Oe/Be stars in these environments.  

\section{Discussion}

The VDD model requires rapid rotation, which stellar winds, and hence high metallicity, inhibit.  Trends with rotational velocity should correlate with Oe/Be star frequencies and our sample of field SMC Oe stars strongly supports the qualitative predictions of this model.  In the metal-poor SMC, the Oe/Be phenomenon should extend to earlier spectral types, and our data confirm this trend, although we caution that the Galactic distribution of Oe star spectral types is dominated by small number statistics.  We find four Oe stars with spectral types earlier than the earliest MW Oe star, and five additional Ope stars which may have temperatures representative of similar, and even earlier, spectral types.  We also measure an Oe/O ratio of $0.26\pm0.04$, which is much greater than the MW value, $0.03\pm0.01$, and is consistent with the expected anticorrelation with metallicity.  Additionally, the pattern in rotation rates between field and cluster environments in the SMC is consistent with the pattern in Oe star frequencies.  Thus, our data show three trends consistent with expectations from the decretion disk model, and confirm that Oe stars are the high-mass extension of the Be phenomenon.  By the same token, our Oe star frequency offers further evidence against the previously proposed wind-compressed disk and magnetic-compressed disk models for the Be phenomenon, because both models rely on strong stellar winds to form a circumstellar disk \citep[e.g.][]{bjo93,cas02}, and therefore predict a correlation with metallicity; this contradicts our observed anti-correlation.

\section{Conclusions}

We present a complete sample of Oe stars identified from RIOTS4, a spatially complete, spectroscopic survey of SMC field OB stars \citep{lamin}.  We find $29\pm1$ Oe stars, of which only two have previously been identified.  Ten are Ope stars, which exhibit \hei\ infill and/or emission.  This complete sample of Oe stars offers an unprecedented opportunity to examine the classical Oe/Be phenomenon in a low-metallicity, field environment.  In particular, our conventional understanding of Oe/Be stars is that they host rotationally induced decretion disks, whose formation is suppressed at higher metallicity due to strong stellar winds that strip angular momentum from the stars.

Thus, weaker stellar winds in the low-metallicity SMC should produce earlier-type Oe stars than in the Galaxy, and this is consistent with our observations.  Our spectral type distribution (Figure~\ref{f:dist}) shows that 21--27\% of our Oe stars are earlier than O7.5e, the earliest MW Oe star spectral type \citep{neg04}.  We find four Oe stars (Figure~\ref{f:earlyOe}) that have spectral types in this range: stars 14324 (O6~V((f))$\rm e_{\rm 2}$), 15271 (O6~III((f))$\rm e_{\rm 1}$), 69460 (O6.5~III((f))e$_2$), and 52363 (O7~III(f)e$_1$).  The O6e stars are the earliest classified Oe stars in the SMC, and possibly the earliest unambiguous, non-Ope classifications of classical Oe stars known to date.  Furthermore, we identify five Ope stars (Figures~\ref{f:earlyOpe} and \ref{f:earlyOpec}) with $\teff$ corresponding to similarly early spectral types.  The hottest Ope star, 77616 (Figure~\ref{f:earlyOpec}), shows He~\textsc{ii}~$\lambda$4686 emission, demonstrating that in low-metallicity environments, even the very hottest O stars can presumably form decretion disks.

In addition, the weak stellar winds in the metal-poor SMC should produce a higher frequency of Oe stars than in the higher-metallicity MW.  Consistent with this prediction, we measure an Oe/O ratio of $0.26\pm0.04$, an order of magnitude greater than our MW estimate, $0.03\pm0.01$.  Furthermore, this trend is also seen in Be stars, which are 3--5 times more frequent among early-type stars in the SMC than in the MW \citep{mar10}.

The correlation between rotation velocities and Oe star frequencies also supports the decretion disk model.  In the Galaxy, field OB stars rotate more slowly than their cluster counterparts, so we expect the Oe/O ratio to be higher in clusters than in the field.  However, in the SMC, no significant difference is observed between field and cluster B star rotation velocities \citep{mar07}.  The Oe/O ratio ($0.25\pm0.05$), as measured for O8--9 stars among our SMC field sample, and the corresponding cluster estimate from \citet{mar10} ($0.24\pm0.09$) are in agreement.  Thus, our data for SMC Oe stars are consistent with predictions for how metallicity, stellar winds, and environment affect decretion disk formation, and confirm that Oe stars are the high-mass extension of the Be phenomenon. 

Another fundamental result to emerge from our study is that for mid and early O spectral types, the Oe/Be phenomenon manifests as Ope spectra, exhibiting \hei\ infill/emission.  This is seen from Table~\ref{t:Opetab}, which illustrates that Ope stars are more common among earlier-type stars.  Additionally, our data suggest that Ope stars are associated with \citet{les68} $\rm Oe_{\rm 3}$ and Oe$_{\rm 4}$ classifications and dominate for spectral types earlier than $\sim$O7.5e.  Furthermore, our earliest ordinary Oe stars are $\rm Oe_{\rm 1}$ or $\rm Oe_{\rm 2}$ stars.  This indicates that photospheric \hei\ and \heii\ absorption lines are visible when disk emission is weak, corresponding to \citet{les68} classifications of Oe$_1$ and Oe$_2$.  The \hei\ disk emission is thus simply coincident with H Balmer emission in the early types.  At the earliest extreme, we even find \heii\ disk emission.  Additionally, Fe~\textsc{ii} emission is more common among $\rm Oe_{\rm 3}$ and Oe$_4$ classifications; approximately 50\% of late-Oe and early-Be type stars have Fe~\textsc{ii} emission.  

We also revise a NIR color selection criterion for Oe stars \citep{bon10} to $J-[3.6]\geq0.1$, which more accurately recovers our spectroscopically identified Oe stars, especially those with weaker Balmer emission (Figure~\ref{f:CMD}).  In addition, we find that Oe stars are systematically brighter than ordinary O stars by about 1 mag in the $J$-band, likely due to either disk emission and/or Oe stars being somewhat evolved. 

Finally, our discovery of the extreme Ope star, 77616, which shows the Oe/Be phenomenon in both \hei\ and \heii\ emission, offers observational support for theoretical predictions that the hottest, rapidly rotating, low-metallicity O stars can reach \heii-producing $\teff$.  The \heii\ Ope emission of star 77616 is consistent with models by \citet{bro11} for stars $\gtrsim30\ \rm M_\odot$ with rotation velocities $> 400$ km/s at SMC metallicity, yielding rotationally mixed stars with $\teff\sim 60,000$ K.  Since Oe stars are the most rapidly rotating O stars, we suggest that extreme Oe/Ope stars may be responsible for He~\textsc{ii} emission occasionally seen in H~\textsc{ii} regions and extreme starburst galaxies. Thus, our results show that even the hottest O stars can present the Oe/Be phenomenon and produce decretion disks.     

\acknowledgements We thank the the referee, Ignacio Negueruela, for useful and insightful feedback.  We also thank Jon Bjorkman, Karen Bjorkman, Xiao Che, Peter Conti, Selma de Mink, Jes\'{u}s Ma\'{i}z-Apell\'{a}niz, Philip Massey, Megan Reiter, John Monnier, Thomas Rivinius, and Aaron Sigut for useful discussions.  We also thank Ian Roederer for helpful comments on a draft of this paper.  Additionally, JBGM thanks Daniel Gifford for a variety of help, including with python programming and statistics.  This work was supported by NSF grants AST-0907758 and AST-1514838.

\bibliography{Oepaper_JGM_acceptedversion}

\clearpage

\appendix

Here we present all of our RIOTS4 Oe and Ope spectra.  We organize the stars by spectral type, from earliest to latest. Figures~\ref{f:Appendix1} and \ref{f:Appendix2} present luminosity class V stars and luminosity class III stars, respectively.  We label the principal absorption lines used for our classification criteria as well as Fe~\textsc{ii}, He~\textsc{i}, and He~\textsc{ii} emission lines. 

\begin{figure*}
\epsscale{1.05}
\plotone{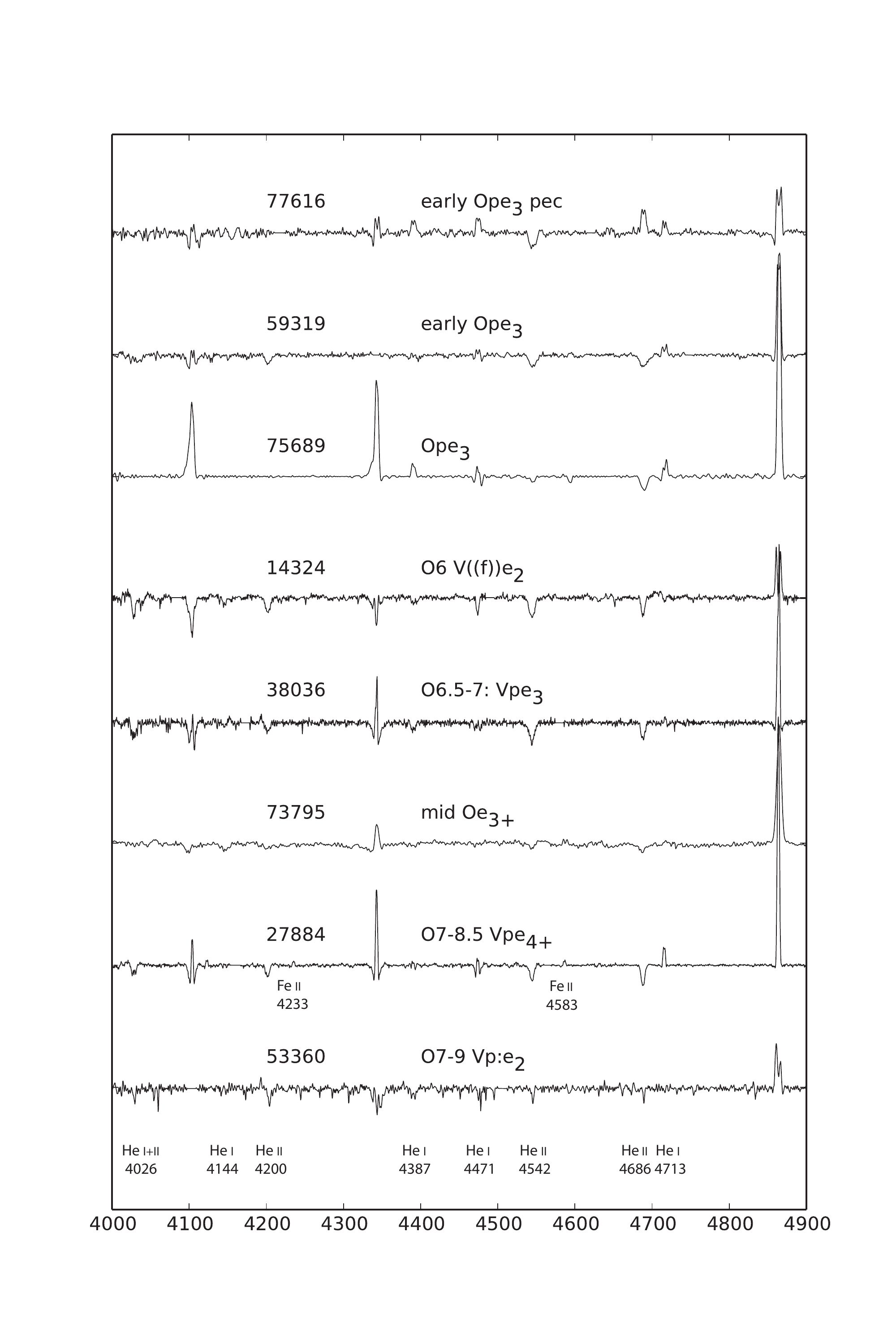}
\caption{Oe/Ope stars with luminosity class V.  The \citet{mas02} ID numbers and spectral classifications are presented above each spectrum.  The absorption lines used to classify these stars are labeled, as well as any He~\textsc{i} and Fe~\textsc{ii} emission lines.} 
\label{f:Appendix1}
\end{figure*} 

\setcounter{figure}{6}
\begin{figure*}
\epsscale{1.05}
\plotone{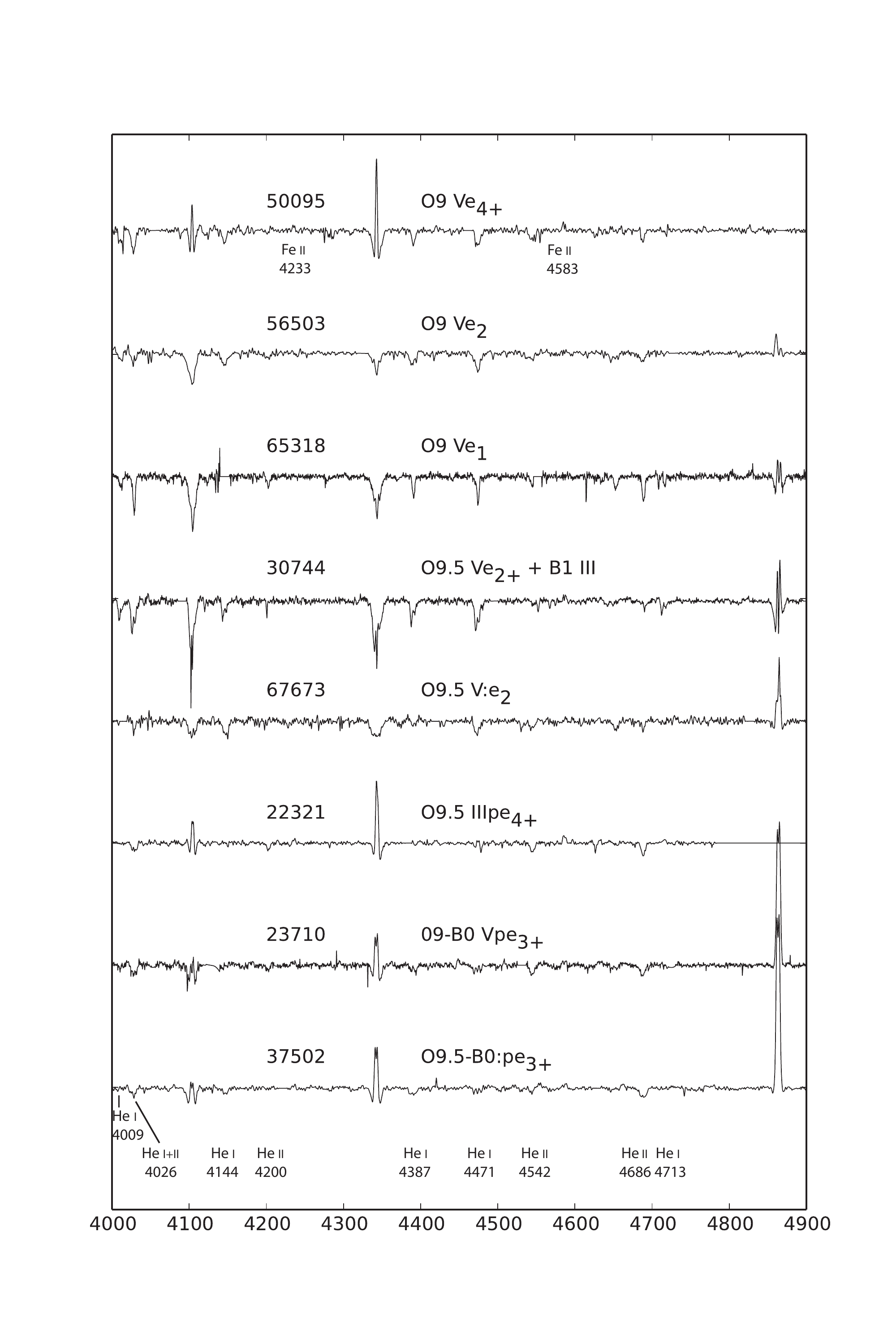}
\caption{continued.}
\end{figure*} 

\begin{figure*}
\epsscale{1.05}
\plotone{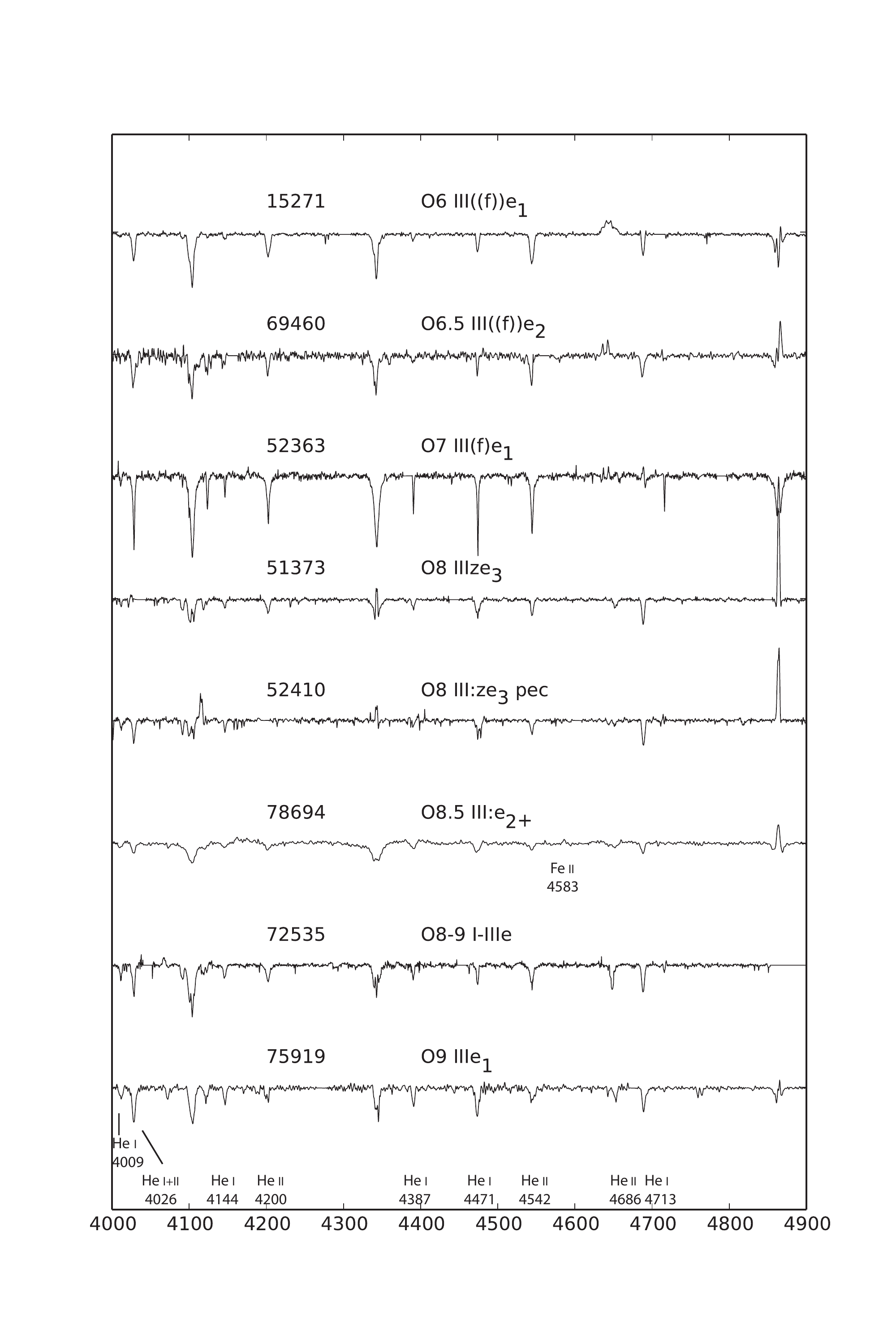}
\caption{Oe/Ope stars with luminosity class III.  Star 52410 is peculiar due to the emission feature near H$\delta$, and star 18329 is peculiar because of the emission feature located near H$\gamma$.  The \citet{mas02} ID numbers and our spectral types appear above each spectrum.  Absorption features used in classifying these spectra and He~\textsc{i} emission lines are labeled.}  
\label{f:Appendix2}
\end{figure*}
\clearpage 

\setcounter{figure}{7}
\begin{figure*}
\epsscale{1.11}
\plotone{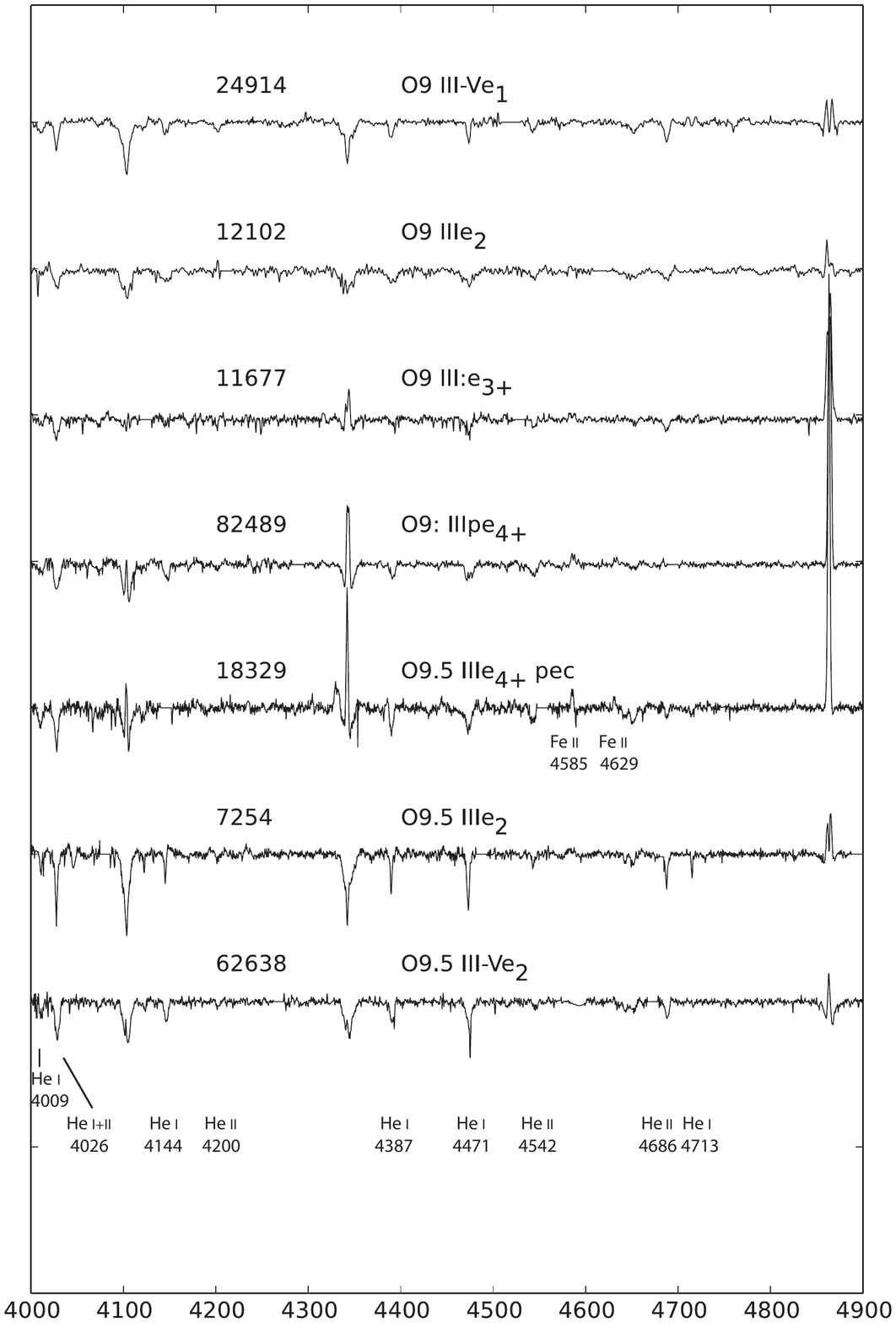}
\caption{continued.}
\end{figure*}

\end{document}